\documentclass[aps,pre,twocolumn,showpacs,superscriptaddress,footinbib]{revtex4}

\usepackage{psfrag,amsmath,amssymb}
\usepackage{epsfig,rotating}
\usepackage{multirow}
\usepackage{graphicx}

\begin{document}

%\preprint{}

%Title of paper
\title{Spin models for  orientational ordering of  colloidal molecular crystals}

\author{Andreja \v Sarlah}
\affiliation{Faculty of Mathematics and Physics, 
Department of Physics, Univerza v  Ljubljani, Jadranska 19, 
SI-1000 Ljubljana, Slovenia}

\author{Erwin Frey} 
\author{Thomas Franosch}
\affiliation{Arnold Sommerfeld Center for Theoretical Physics (ASC)
and Center for NanoScience (CeNS), Department
of Physics, Ludwig-Maximilians-Universit{\"a}t M{\"u}nchen,
Theresienstr. 37, D-80333 M{\"u}nchen, Germany}

\date{\today}

\begin{abstract}
Two-dimensional colloidal suspensions exposed to periodic external
fields exhibit a variety of molecular crystalline phases. There two or more 
colloids assemble at lattice sites of
potential minima to build new structural entities, referred to 
as molecules. Using the strength of the potential and the filling fraction 
as control parameter, phase transition to unconventional orientationally
ordered states can be induced. We introduce an approach that focuses at 
the discrete set of orientational states relevant for the phase
ordering. The orientationally ordered states are mapped to
classical spin systems. We construct effective hamiltonians for
dimeric and trimeric molecules on triangular lattices  suitable
for a statistical mechanics discussion. A mean-field analysis
produces a rich phase behavior which is substantiated by Monte
Carlo simulations.
\end{abstract}


\maketitle

\section{Introduction}
Soft materials comprised of colloidal particles undergo phase
transitions from fluid to crystalline order just as their atomic
counterparts~\cite{chaikin1995,pusey1986,lowen1994,russel1995}.
The intrinsic time and length
scales of such suspensions offer the advantage to monitor the
ordering phenomena by video microscopy on a single particle
level. Furthermore, the interactions can be tailored to a certain
degree by controlling the surface charge of the colloids or
changing the refractive index of the particles and the solvent or
by screening electrostatic interaction by the addition of salt,
etc.~\cite{russel1995,israelachvili1985,likos2001}.
Colloids thus constitute fascinating model
systems to study material properties. For example, 
new crystalline structures
without known atomic analogue have been found~\cite{leunissen2005} for
mixtures of charged colloids.

In particular, confining geometries allow to study phase
transitions in two-dimensional systems where strong thermal
fluctuations often become  the  relevant mechanism   for the
melting of order. For example, close
to melting a universal relation betwen the elastic constants 
is predicted  in the framework of the
Kosterlitz-Thouless-Halperin-Nelson-Young
theory~\cite{kosterlitz1973,nelson1979,nelson2002} as has been recently
observed experimentally~\cite{zahn1999}. Even richer phase
behavior is expected if quasi two-dimensional colloidal systems
are exposed to external potentials as can be realized by
interfering laser
beams~\cite{chowdhury1985,wei1998,bechinger2001-1}, periodic
pinning arrays~\cite{mangold2003}, imprint or stamping
techniques~\cite{lin2000}. For example, in the case of
one-dimensional troughs  exotic phases such as the locked smectic
phase or a floating solid have been
predicted~\cite{frey1999,radzihovsky2001}  and partially observed
in simulations~\cite{enomoto2005,reichhardt2005} and
experiments~\cite{baumgartl2004}.

For three or more laser beams periodic patterns with two
modulation directions can be achieved~\cite{brunner2002}, in
particular, square and triangular lattices have been
studied~\cite{brunner2002,reichhardt2002,agra2004,reichhardt2005-1}.
The colloidal particles  then accumulate in the potential minima
thus imposing a modulation of the colloid density. The competition
between the attraction to the potential minima and the mutual
repulsion of the particles can be conveniently controlled by
adjusting the {\em filling factor}, i.e., the number of colloidal
particles per potential minimum. For instance, by changing the
angles of incidence the lattice constant of the periodic
modulation can be varied continuously. Since the electrostatic
interaction is usually screened by counterions in the solvent, the
lattice spacing also sensitively determines how strongly particles
interact. A second route to affect the balance between external
compression  and internal repulsion is by changing the intensities
of the laser beams. Third the salt concentration of the solvent is
a suitable control parameter to tune  the
range of the repulsive interaction, i.e. the Debye screening
length.

\begin{figure}[htbp]
\includegraphics[width=1.0\linewidth]{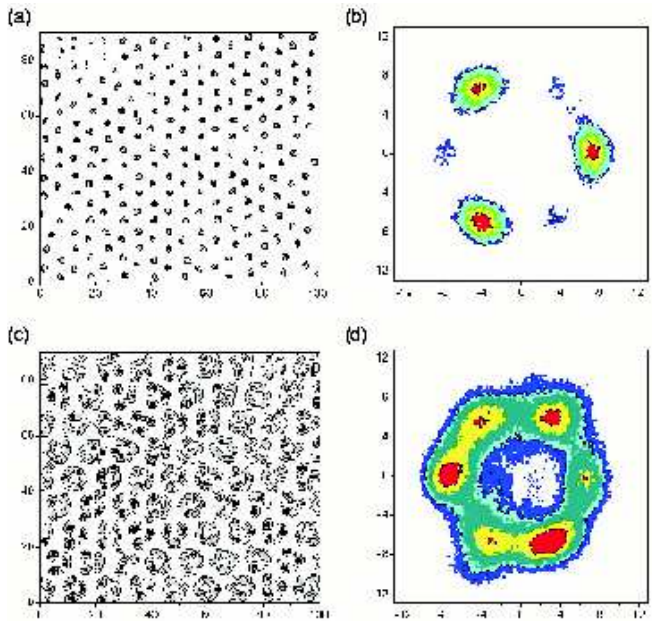}
\caption{Contour plots of the lateral density distribution (left) 
and the averaged local particle density (right) for different 
light potentials: $V_0=60~k_B T$ (top) and $V_0=110~k_B T$ (bottom). 
In the first case the majority of three-groups of colloids is 
oriented in the same way, whereas in the later case the orientation 
of the triangles is uncorrelated at large distances. At some 
intermediate $V_0$ a phase transition occurs.
The horizontal/vertical axes are $x$ and $y$, respectively. All 
units are in $\mu$m. Figure is reproduced from a part of Figure~2 of 
Brunner and Bechinger Ref.~\cite{brunner2002}.}
\label{fig-0-1}
\end{figure}

This paper is motivated by recent experimental studies on
two-dimensional colloidal system exposed to a potential of a
triangular lattice symmetry with a filling factor
three~\cite{brunner2002}. The density of the colloids has been
 selected in a regime below
the spontaneous freezing transition. 
A continuous increase of the strength of
the laser field has induced  a sequence of qualitative changes of
order. 
At low intensities of the
lasers, corresponding to a potential depth of say $V_0 \sim k_B
T$, the colloids exhibit the typical response of a fluid to an
external modulation, i.e., a small periodic component is
superimposed to the homogeneous background density. Enhancing the
strength of the laser potential by an order of magnitude the
response becomes highly nonlinear; groups of three colloidal
particles cluster in the close vicinity of each potential minimum.
However, a significant number of {\em defects} -- groups of two or
four colloidal particles -- is still present in the system. Since
the filling factor has been adjusted to three, the number of
four-groups equals the one of two-groups. In this regime the
barriers separating minima become high and correspondingly single
particle diffusion is strongly suppressed~\cite{reichhardt2002},
similar to a solid. Although no thermodynamic phase transition has occurred the
system is referred to as a {\em molecular crystal}. A further
increase of the external potential results in a freeze-out of the
defects, i.e., each lattice site is occupied by precisely three
colloids. The confinement is rather strong, balanced only by the
mutual electrostatic repulsion, leading to triangular arrangements
of the three-groups aligned with the lattice. Interestingly, the
majority of three-groups is oriented in the same way with a small
number of groups adopting a mirrored configuration, see 
Fig.~\ref{fig-0-1}. Using a suitable order parameter, 
orientational correlations demonstrate the
long-range orientational order. At still higher laser potentials
the orientation of the triangles is uncorrelated at large
distances, indicating a phase transition to an orientationally
disordered phase. In Ref.~\cite{brunner2002} this subsequent
melting has been referred to as reentrant transition, see 
Fig.~\ref{fig-0-1}.

The possibility to induce  phase transitions by periodic laser
patterns has stimulated theoretical studies in two-dimensional
colloidal systems.  
Reichhardt
and Olson have performed extensive Langevin simulations 
and have identified ground states for a series of filling
factors on square and triangular lattices~\cite{reichhardt2002}.
In particular, they have also observed the orientational ordered state
corresponding to the experiment discussed above. For non-integer
filling factors the  molecular crystals are composed of
supercells, i.e., several types of groups of colloids are distributed
periodically on the optical lattice~\cite{reichhardt2005}. By
lumping particles to rigid molecules, explicit minimizations of
intermolecular interactions in the presence of the optical
potential have revealed several ordered ground
states~\cite{agra2004}. Furthermore, an energy functional for
dimeric molecules on a square lattice has been introduced and an
analogy to Ising behavior has been identified~\cite{agra2004}.

The goal of the present paper is to elaborate a theoretical
framework for colloidal molecular crystals in external fields
built from a statistical mechanics perspective, as outlined
in~\cite{sarlah2005}. First, to deal with the complexity of the
system we focus on the essential low-energy degrees of freedom of
colloidal suspension. From the experimental observation one infers
that the groups of colloids are to be considered as rather rigid
entities as soon as the laser potential exceeds several tens of
$k_B T$. The entities are referred to as molecules or dimers,
trimers, etc. The key idea is to keep only discrete orientational
states of the molecule as fluctuating variables, i.e., relevant
for a statistical mechanics model. The separation of energy
scales, e.g. the condensation or binding energy to build such
molecules is much larger than the thermal scale, suggests to
ignore processes connected with the breaking apart of the
structure. The
theoretical description we wish to develop is of
phenomenological nature, since it does not explain the formation of
molecules. In particular, the regime of low laser intensities
where the molecular crystal exhibits a significant fraction of
defects is not within the scope of our approach. The benefit of
focusing on effective low-energy excitations is that one can
derive simple models that are amenable to powerful methods of the
statistical mechanics toolbox. In particular, it allows to
identify the appropriate broken symmetry phases and to map the
problem to  related magnetic transitions for which a great wealth
of knowledge has been developed.

Specifically, in Section II we develop a theory for the experiment
performed by Brunner and Bechinger~\cite{brunner2002}. We identify
the relevant excitations of the system at the thermal scale as
single trimer flips. A hamiltonian is constructed that couples
nearest neighboring trimers via their respective orientations. The
microscopic interactions, that are for example also responsible
for the formation of the molecules, enter the theory only
implicitly through three energy scales. The explicit link to the
experimental parameters, e.g., salt concentration, effective
surface charge, etc., is deferred to the Appendix. It turns out
that the statistical mechanics of trimers on a triangular lattice
maps to an Ising model once the spectrum of single trimer
excitations are evaluated. A comparison to experimental results
corroborates our approach.

In Section III we propose an experimental setup using a
triangular lattice with a filling factor two. Then dimers are the
composite objects and three orientational states per lattice site
emerge. The corresponding hamiltonian can be understood as a
natural extension of a three-states Potts model with the
peculiarity that orientational pair interactions depend on the
lattice orientation. We find a rich scenario of four broken
symmetry phases occupying different regions in the phase diagram.
In addition to ``ferro-'' and ``antiferromagnetic'' structures a
herring-bone structure appears and a novel phase referred to as
{\em Japanese 6 in 1} (J6/1) is discovered. An analytical approach
based on a variational mean-field calculation is presented to
explore the phase boundaries. Extensive Monte Carlo simulations
are employed to corroborate our findings and we present a careful
discussion of the different phases.

In the Conclusion, we summarize our main results and put them in a
broader perspective of experimental and theoretical solid state
physics.

\section{Trimers}
\begin{figure}[htbp]
\includegraphics[width=1.0\linewidth]{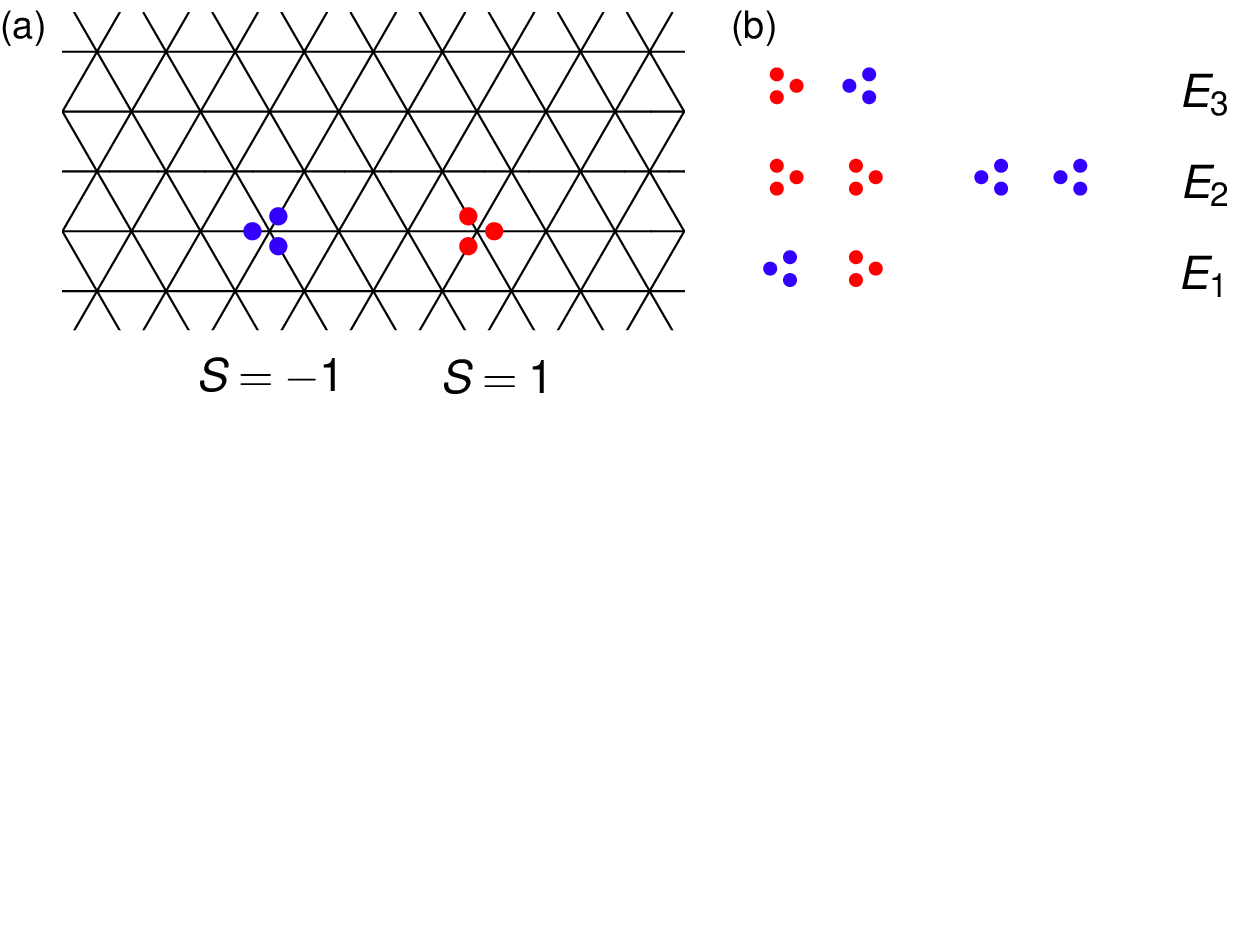}
\vspace{-4.25cm}
\caption{The model system. (a) The triangular lattice of the external 
field and the two discrete orientational states of the trimers. 
(b) Three interaction energies  for the orientational configurations
of neighboring trimers.}
\label{fig-2-1}
\end{figure}

For  densities of the colloidal particles  adjusted  such  that
there are three particles per potential minimum, trimers will form
spontaneously at sufficiently strong laser intensity. Then the
low-energy effective degrees of freedom are given by the discrete
orientational states of the composite object. Due to the symmetry
of the optical lattice the trimers constitute equilateral
triangles which are aligned with the triangular lattice. From
Fig.~\ref{fig-2-1} one infers two energetically equivalent
orientational states of the trimer corresponding to the optimal
balance between the compression of the constituent colloids due to
the laser potential and their mutual screened Coulomb repulsion.
The residual interaction between trimers depends on their relative
orientation as well as on their relative lattice position. Since
the residual interaction is solely due to a screened Coulomb
interaction the magnitude of this energy decreases rapidly with
increasing distance. For typical experimental setups the
screening length is an order of magnitude smaller than the lattice
constant \cite{brunner2002}, 
which allows to restrict the residual interaction to
nearest neighbors. For each neighboring pair of trimers there are
four possible geometrical configurations, two of which are mirror
images of each other, see Fig.~\ref{fig-2-1}. These configurations
introduce three energy scales, with generic ordering $E_1 < E_2 <
E_3$. These scales are functions of the screening length, the
strength of the laser potential, as well as the effective charges
of the constituent colloids. Appendix~\ref{appendix-1} provides a
derivation of the explicit relation and supplements a careful
discussion of the concept of composite objects.

\begin{figure}[htbp]
%\vspace{-0.45cm}
\includegraphics[width=1.0\linewidth]{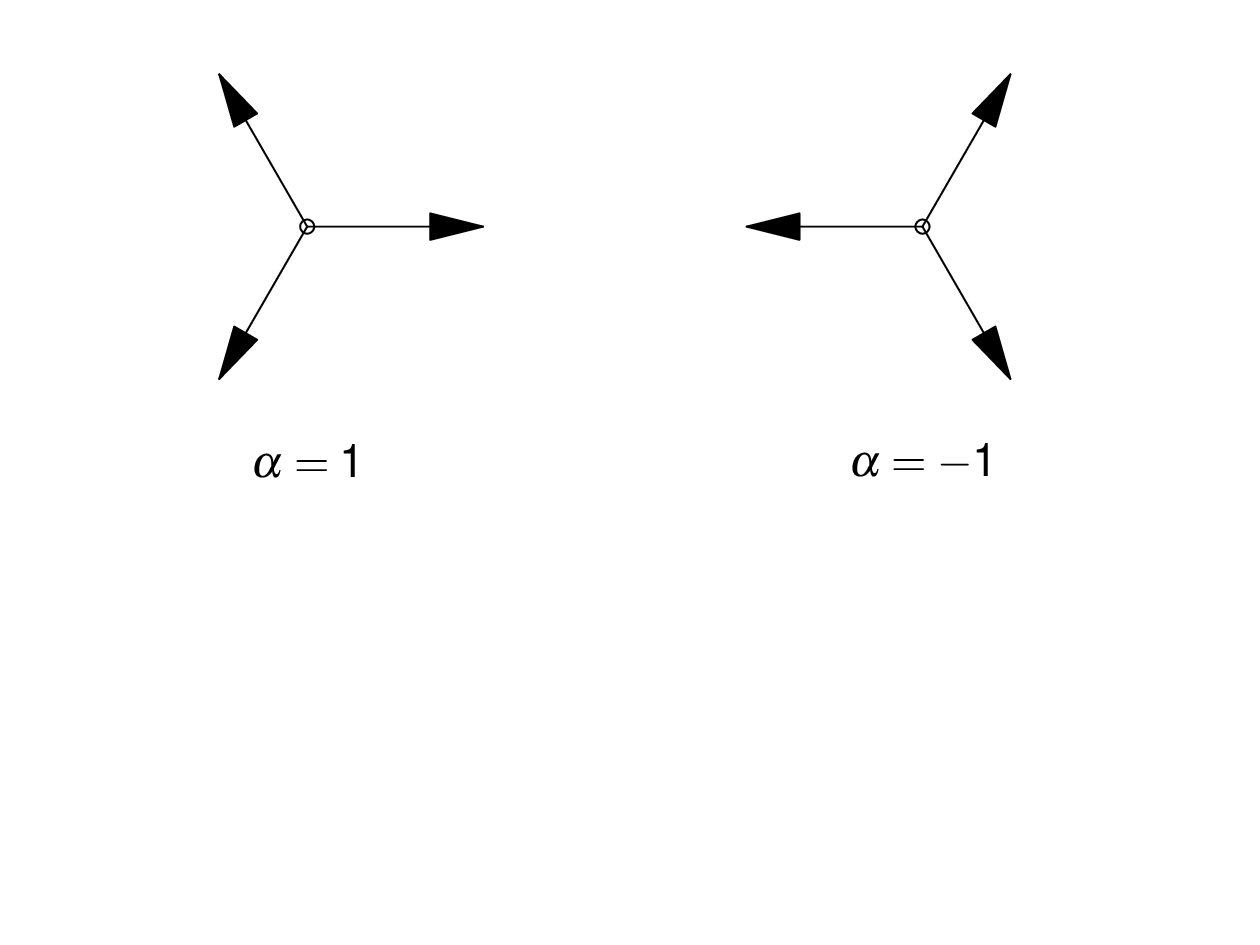}
%\vspace{-2.5cm}
\vspace{-3.5cm}
\caption{The two  classes of equivalent bond vectors in the 
trimer system. Note that the classes have the same symmetry as 
the corresponding trimer state, i.e., for $S_i = \alpha$.}
\label{fig-2-2}
\end{figure}

The considerations of the previous paragraphs allow to define the
model in terms of a statistical mechanics problem. Each trimer $i$
is attributed a ``spin state'' $S_i = \pm 1$, the total number of
orientational configurations being given by $2^N$ for a triangular
lattice of $N$ sites. The phase behavior of the trimer system is
encoded in a ``spin hamiltonian'', which reduces to a sum over
local interaction energies. The trimer system exhibits a
peculiarity in the sense that these energies are dependent on the
{\em bond vector}, i.e., the direction of the connecting lattice
vector, in addition to the two spin orientations of the
neighboring spins. Explicitly, a simultaneous flip of both spin is
not a symmetry operation of the hamiltonian, however, an
additional rotation of the corresponding bond vector by $60^\circ$
restores the original configuration. Consequently, a rotation of
the bond vector by $120^\circ$ with identical spin states results
in the same interaction energy. Therefore, similar to the two
orientational states of a trimer, there are only two inequivalent
bond vectors, see Fig.~\ref{fig-2-2}. The hamiltonian can be
expressed in the following form,
\begin{eqnarray}
    {\cal H}  = \sum_\alpha \sum_{\langle ij \rangle_\alpha}
        h(S_i,S_j;\alpha) ,
\label{eqn-2-1}
\end{eqnarray}
where $\alpha=\pm 1$ denotes the orientation of the bond vector,
and the symbol $\langle ij \rangle_\alpha$ indicates the
neighboring pair of spins with directed bond vector $\alpha$. Hence, 
each bond appears precisely once in the entire sum.  
Reversing the order of the lattice sites implies an inversion of
the bond vector $\langle ij \rangle_{\alpha} = \langle ji
\rangle_{-\alpha}$. Explicitly, the local energy functionals read
\begin{eqnarray}
\parbox{3cm}{
\begin{center}
    $h(S_i,S_j;1)$ \\[0.1cm]
    $\begin{array}{|l|cc|}
        \hline
        {}_{S_i}\!\! \left\backslash \!{}^{S_j} \!\! \right. & 1 & -1 \\
        \hline
        \ \ 1 & E_2 & E_3 \\

        -1 & E_1 & E_2 \\
        \hline
    \end{array}$
\end{center}
}
\parbox{3cm}{
\begin{center}
    $h(S_i,S_j;-1)$ \\[0.1cm]
    $\begin{array}{|l|cc|}
        \hline
        {}_{S_i}\!\! \left\backslash \!{}^{S_j} \!\! \right. & 1 & -1 \\
        \hline
        \ \ 1 & E_2 & E_1 \\
        -1 & E_3 & E_2 \\
        \hline
    \end{array}$
\end{center}
}
\label{eqn-2-2}
\end{eqnarray}
Note that due to the above explained dependence on the orientation
of the bond vector, the corresponding energy ``matrices'' are not
symmetric.

\begin{figure}[htbp]
\includegraphics[width=1.0\linewidth]{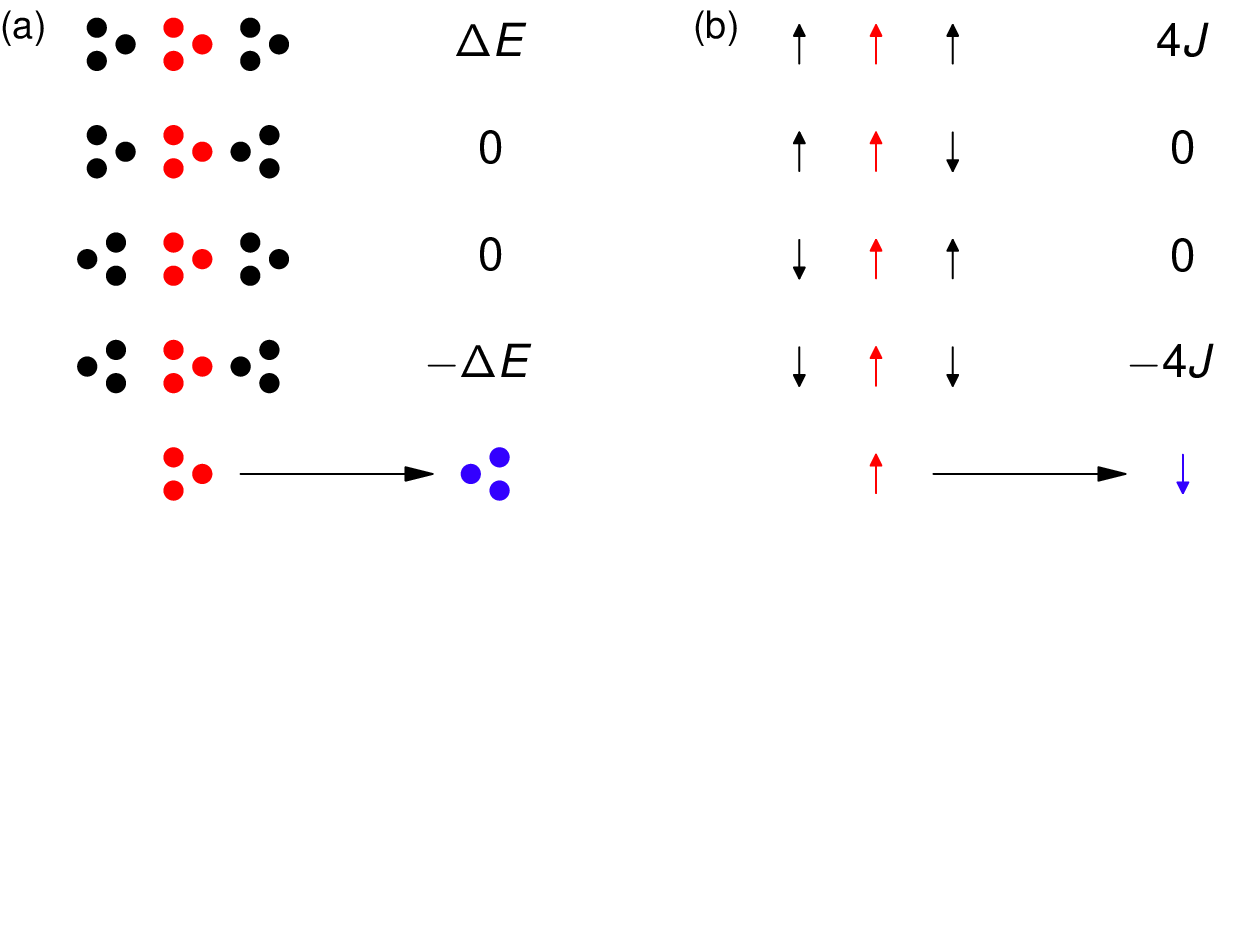}
\vspace{-3.5cm}
\caption{(a) Schematic representation of all possible situations for
a trimer flip with the corresponding energy changes; 
$\Delta E = E_1 + E_3 - 2 E_2$. (b) The same as in (a) but for spins
of the spin-$1/2$ Ising hamiltonian
${\cal H} = -J \sum_{\langle ij \rangle} S_i S_j$.}
\label{fig-2-3}
\end{figure}

A global shift of the energy scale does not affect the phase
diagram, leaving two independent energy differences. Introducing
the thermal scale $k_B T$ allows to construct two dimensionless
parameters, and na{\"i}vely one would expect a two-dimensional phase
diagram. However, we shall show now that due to geometrical
constraints there is a one-to-one correspondence of the excitation spectrum
of the colloidal trimers and an Ising model on  a triangular lattice.
Since the interaction energy is additive, 
it is sufficient to consider the case of three
collinear  trimers --- i.e., considering the smallest neighborhood
of trimers with representatives of both inequivalent bond
orientations. If we fix the central trimer there are $2^2$
possible configurations of the two neighbors, all of them are
depicted in Fig.~\ref{fig-2-3}~(a). Then, flipping the central
trimer yields the energy changes $\pm \Delta E,\, 0$ where we have
defined $\Delta E\equiv E_1 + E_3 - 2 E_2$. Although unexpected
due to the bond-oriented local interaction energies, the trimer
system reduces to an Ising model, i.e., with spins allowing for
two distinct orientations and a single energy scale $\Delta E$ for
excitations, see Fig.~\ref{fig-2-3}~(b). As a result of these 
considerations, the hamiltonian
of Eq.~(\ref{eqn-2-1}) is equivalent to
\begin{equation}
    {\cal H} = - J \sum_{\langle ij \rangle} S_i S_j + {\cal H}_0,
\label{eqn-2-3}
\end{equation}
where $\sum_{\langle ij \rangle}$ indicates the sum over all pairs of
neighboring spins, irrespective of the orientation of the pair with respect to
the lattice. The exchange energy $J$ sets the scale for local
excitations and is related to the three interaction energies by
\begin{equation}
    J = {1 \over 4} \Delta E = {1 \over 4} (E_1 - 2 E_2 + E_3) .
\label{eqn-2-4}
\end{equation}

The additive constant can be evaluated to ${\cal H}_0 = - \frac{3}{4}
N (E_1 + 2 E_2 + E_3)$. Although the energy scales possess the
generic ordering $E_1 < E_2 < E_3$, the sign of the exchange
coupling $J$ can attain in principle both negative and positive
values. For two-dimensional colloidal systems interacting via
strongly screened Coulomb interaction explicit evaluation yields
positive values, i.e., ''ferromagnetic'' coupling.
For systems with different microscopic interactions, e.g., with 
paramagnetic particles, an ``anti-ferromagnetic'' phase could possibly be 
realized.

Let us emphasize that the mapping of the trimer hamiltonian to an Ising
model is different from the 1:1 correspondence of a 
lattice gas model for a binary alloy to an Ising system. 
There the nearest-neighbor interaction energy is parametrized also by
three energy scales $\epsilon_{AA}, \epsilon_{AB}, \epsilon_{BB}$ 
corresponding
to a pair of two neighboring $A$ atoms, a mixed pair of an $A$ and a $B$ atom,
and two $B$ atoms. 
For the mixed pair it is irrelevant whether atom $A$ or 
$B$ is left/right or up/down. For the trimer problem the configurations where 
trimers face each other or are back-to-back are energetically 
rather different, see Figure~\ref{fig-2-1}~(b). 

For a two-dimensional Ising model an exact solution is available
for various lattice symmetries~\cite{baxter1989}. On a triangular
lattice the transition between the ``ferromagnetic'' and
``paramagnetic'' phase occurs at $k_B T_c = 4 J / \ln 3$. In the
experiments of Ref.~\cite{brunner2002}, the
temperature is kept fixed and the intensity of the external
potential is varied. This variation is accompanied by the change
of the linear extensions of trimers, see
Appendix~\ref{appendix-1}, which further affects the strength of
the exchange energy $J$. Above some critical laser intensity the
exchange energy becomes smaller than the critical one,
\begin{equation}
    J < J_c = k_B T \, {\ln 3 \over 4},
\label{eqn-2-5}
\end{equation}
and the orientational order of the trimer system is lost.

\begin{figure}[htbp]
\includegraphics[width=1.0\linewidth]{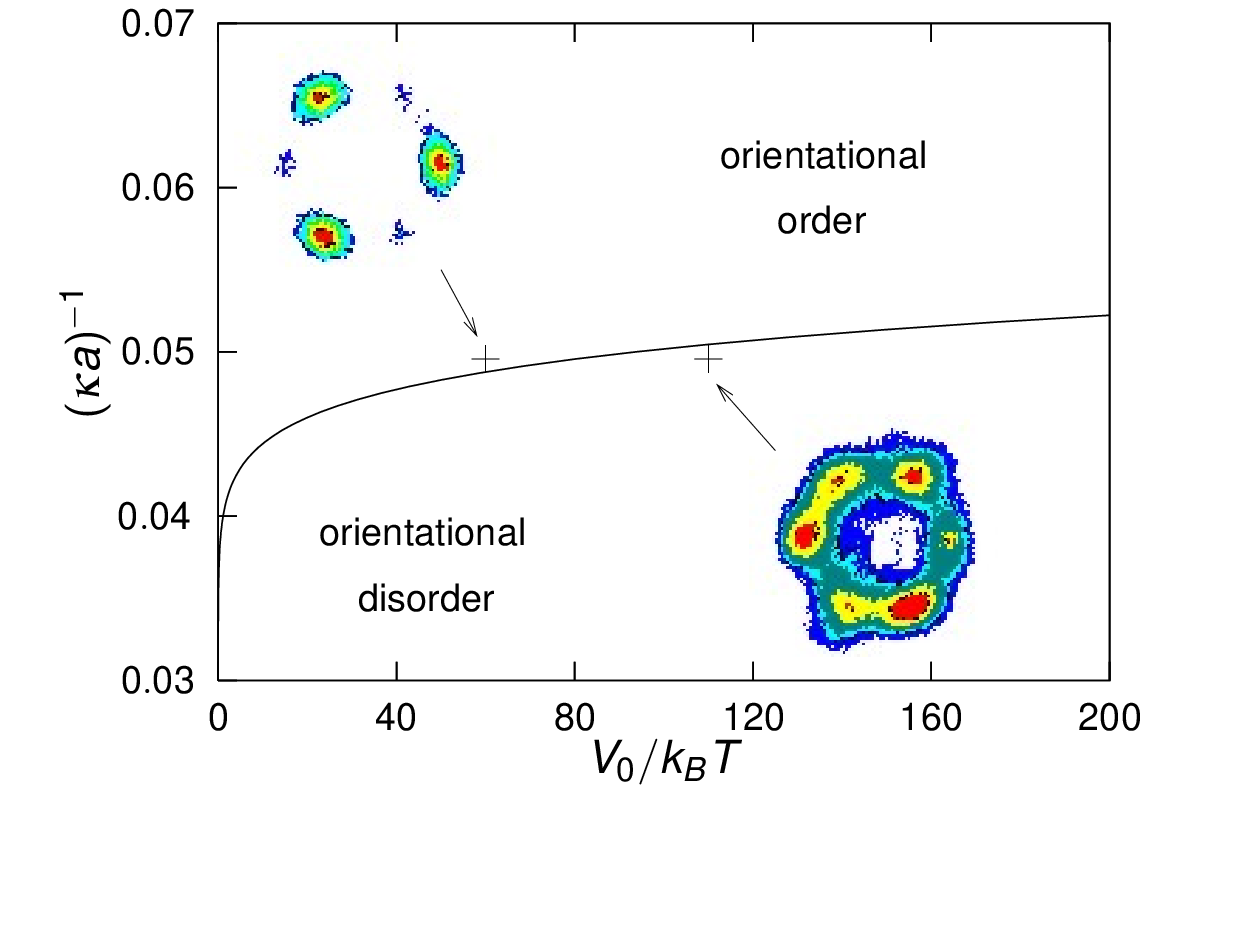}
\vspace{-1.5cm}
\caption{Phase diagram of the colloidal trimer system as a function 
of the strength of the external potential (in units of $k_B T$ where 
$T=295$~K) and colloid-colloid interaction characterized by a product 
of the inverse Debye length and the lattice constant (solid line). 
The two crosses correspond to the experimental observations reported 
in Ref.~\cite{brunner2002}. Insets: averaged local particle densities 
as represented in Figs.~2~(f) and (h) of  Ref.~\cite{brunner2002}.}
\label{fig-2-4}
\end{figure}

Employing experimental parameters from  Ref.~\cite{brunner2002},
$1/\kappa=570$~nm for the inverse screening length, ${\cal
E}_0=3.4\times 10^4~k_B T$ for the strength of the screened
Coulomb interaction, and $T=295$~K as room temperature (see
also Appendix~\ref{appendix-1}) we have determined the phase
diagram in the $(V_0/k_B T,(\kappa a)^{-1})$ plane and compared it
to the experimental results. Brunner and Bechinger report  that
for $V_0=60~k_B T$ the trimer system (realized for a lattice
constant $a=11.5~\mu$m) is orientationally ordered, whereas it is
disordered for $V_0=110~k_B T$. The critical strength of the laser
potential found in our calculations is in striking agreement with
these observations, i.e., $V_{0c}=78.6~k_B T$, see also
Fig.~\ref{fig-2-4}. A more systematic study,
 similar to what has been performed in the case of  1D periodic laser
potentials~\cite{bechinger2001}, would be highly desirable, in
particular, in the regime of larger lattice constants and smaller
screening lengths where the shape of the phase boundary is more
sensitive to control parameters.

\begin{figure}[htbp]
\includegraphics[width=1.0\linewidth]{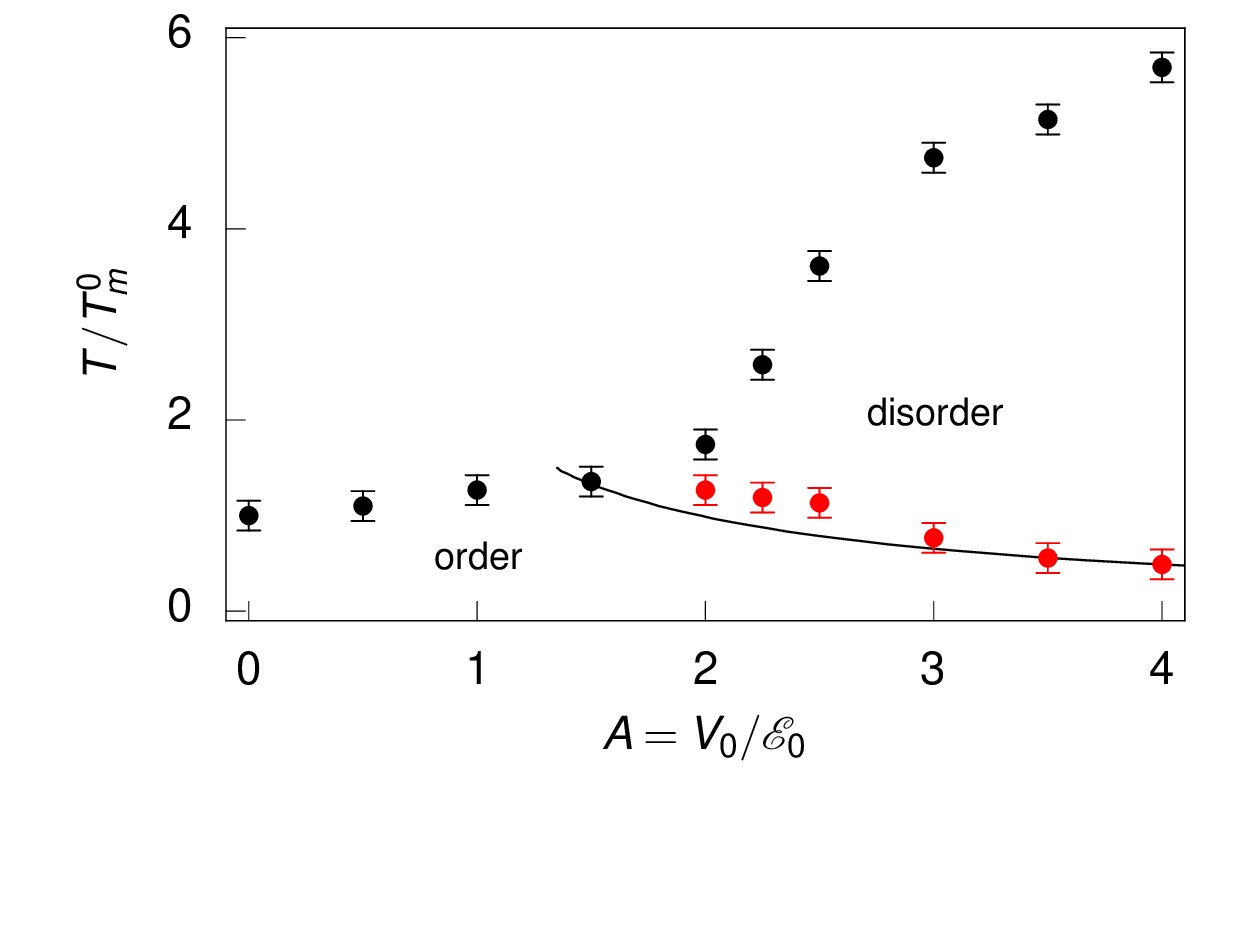}
\vspace{-1.5cm}
\caption{Phase diagram of the colloidal trimer system as a function 
of the reduced strength of the external potential and temperature. Our 
model describes part of the phase diagram corresponding to the 
orientational melting -- solid line. Symbols correspond to the results 
from the simulation of Reichhardt and Olson, see 
Ref.~\cite{reichhardt2002} Fig.~4..}
\label{fig-2-5}
\end{figure}

We have also
compared our theory  to Langevin dynamics simulation results of
Reichhardt and Olson~\cite{reichhardt2002}. The phase diagram is
exhibited in Fig.~\ref{fig-2-5}  in the  $(V_0/{\cal E}_0,T/T_m^0)$ plane.
Although the simulation has been  performed  for dimers on a
square lattice,  the basic mechanism of the orientational melting
-- and thus the corresponding behavior of the phase boundaries --
is robust with respect to the symmetry of the composite objects
and the underlying lattice. Note that the parameter $V_0/{\cal E}_0$ is
by two orders of magnitude smaller than in the corresponding
experimental setup; nevertheless our analytical solution captures
the main features of the phase boundary, i.e. shape and parameter
range. The remaining  phase boundaries presented in the paper of
Reichhardt and Olson are connected to the fission of the composite
objects and are beyond the scope of  our model.

\begin{figure}[htbp]
\includegraphics[width=1.0\linewidth]{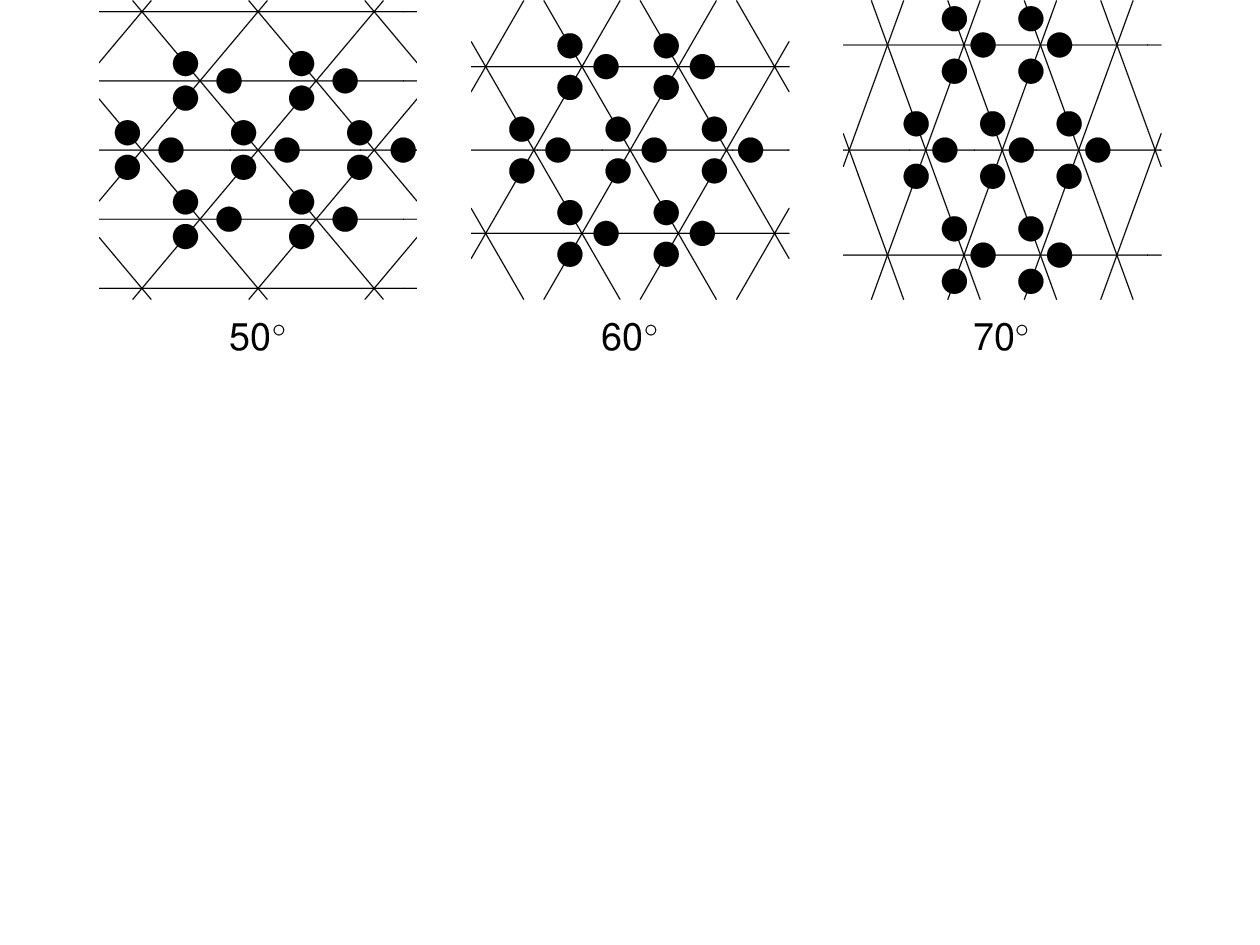}
\vspace{-4.5cm}
\caption{Schematic representation of the lattice with one trimer and
his neighbors for isotropic triangular lattice (middle) and
uniaxially deformed ones; below the angle between the horizontal
bond and the non-horizontal one. Due to the symmetry, the exchange 
coupling in the horizontal direction differs from the one in the
remaining directions.}
\label{fig-2-6}
\end{figure}

Let us comment on the possibility to impose a small uniaxial
strain on the triangular lattice by varying the angles of the
incident laser beams. For small deformations the trimers will
merely stretch leaving the system still with  two orientational
states per lattice site, see Fig.~\ref{fig-2-6}. Since the
distances between pairs of
trimers are now different in different directions, the pair
interactions become direction dependent. The analysis for the
excitations energies is now valid separately for each direction
of the three co-linear trimers. For the direction perpendicular
to the uniaxial deformation the exchange energy differs from
the remaining two, $J_a$ and $J_b$, respectively. The system is
still represented by a two-dimensional Ising hamiltonian, however,
with anisotropic exchange energies. This system again allows for
an exact solution, in particular, the critical temperature is now
determined by the equation~\cite{baxter1989}
\begin{equation}
    \, {\rm e}^{-4 J_a / k_B T} + 2 \, {\rm e}^{- 4 J_b / k_B T} = 1 .
\label{eqn-2-6}
\end{equation}
In conclusion, a small uniaxial strain leads only to a small shift
of the critical temperature but does not lead to any qualitatively
new phase behavior.

\section{Dimers}
\begin{figure}[htbp]
\includegraphics[width=1.0\linewidth]{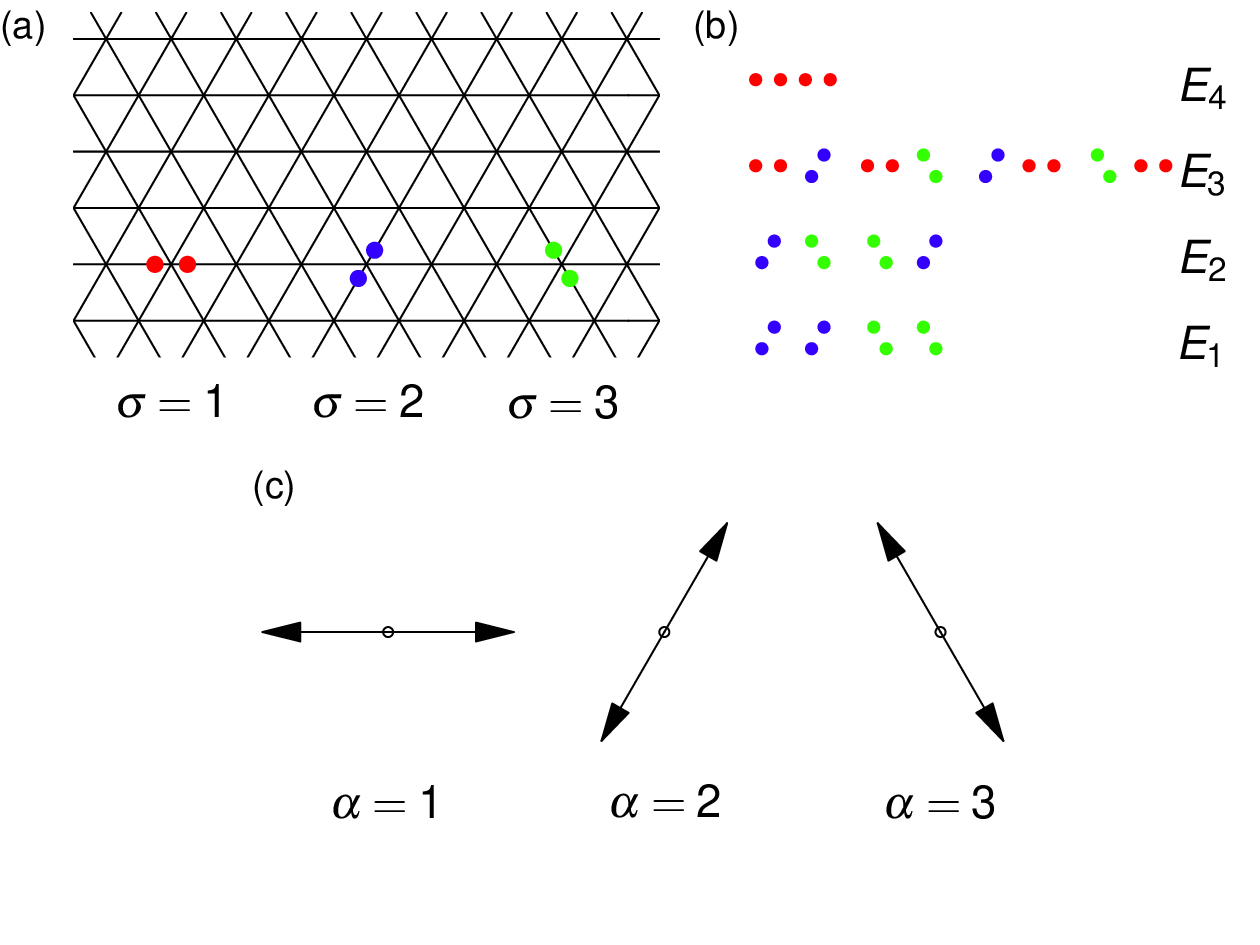}
\vspace{-1.25cm}
\caption{The model system. (a) The triangular lattice of the external
field and the three discrete orientational states of the dimers.
(b) Four interaction energies for the orientational configurations
of neighboring dimers. (c) The three classes of equivalent bond 
vectors in the dimer system. The groups have the same symmetry as the 
corresponding dimer state, i.e., for $\sigma_i = \alpha$.}
\label{fig-3-1}
\end{figure}

In this Section we generalize the concepts introduced for trimers
and discuss the possibility of dimeric molecular crystals and
their corresponding phase diagram.
 Inclining the incident laser beams with respect
to the plane of the two-dimensional colloidal system one can
adjust the lattice constant of the laser potential. Then, by
appropriate matching of the lattice constant to the colloidal
particle density, one can achieve an average of two particles per
potential minimum. For sufficiently large laser intensity, one
expects  that a defect-free structure is formed, i.e., the
composite objects of the system are {\em dimers}. The symmetry of
the lattice allows now for three different orientational states
per lattice site denoted by $\sigma_i = 1,2,3$, see
Fig.~\ref{fig-3-1}~(a).

\subsection{The dimer hamiltonian}
As in the case of trimers, the interaction of neighboring dimers
is the origin of orientational ordering, whereas the balance of
the internal repulsion and the laser pressure merely yields  the
binding energy of the dimers. From Fig.~\ref{fig-3-1}~(b) one
infers that there are four interaction energies for the $3^2$
configurations of neighboring dimers with generic ordering $E_1 <
E_2 < E_3 < E_4$. Explicit formul{\ae} for the relation to
experimental parameters are presented in
Appendix~\ref{appendix-1}. The dependence of the interaction of
dimers on the orientation of the bond vector follows from the
symmetry of the composite objects, i.e., there are three
inequivalent pairs of lattice orientations $\alpha =1,2,3$, 
depicted in Fig.~\ref{fig-3-1}~(c). Since
each pair of equivalent lattice orientations is now collinear,
reversing the order of sites in a directed bond $\langle ij
\rangle_\alpha$ yields a directed bond within the same equivalence
class. This implies that the local interaction energy is symmetric
with respect to exchange of the dimers.

The total hamiltonian of the two-dimensional dimer system
corresponds to a sum over all interactions of neighboring dimer
states. Denoting by $\langle ij \rangle_\alpha$ the pair of
nearest neighbors $i,j$  whose bond vector is parallel to the
orientation of the dimer state $\alpha$, the colloidal dimer
hamiltonian can be expressed in the following form
\begin{equation}
    {\cal H} = \sum_{\alpha = 1}^3 \sum_{\langle ij \rangle_\alpha}
          h(\sigma_i,\sigma_j;\alpha).
\label{eqn-3-1}
\end{equation}
The direction-dependent local energy functionals
$h(\sigma_i,\sigma_j;\alpha)$ are given by:
\begin{eqnarray}
\parbox{8cm}{
\begin{center}
\parbox{2.5cm}{
\begin{center}
    $h(\sigma_i,\sigma_j;1)$ \\[0.1cm]
    $\begin{array}{|l|ccc|}
        \hline
        {}_{\sigma_i}\!\! \left\backslash \!{}^{\sigma_j} \!\! \right. & 1 & 2 & 3 \\
        \hline
        1 & E_4 & E_3 & E_3 \\
        2 & E_3 & E_1 & E_2 \\
        3 & E_3 & E_2 & E_1 \\
        \hline
    \end{array}$
\end{center}
}
\parbox{2.6cm}{
\begin{center}
    $h(\sigma_i,\sigma_j;2)$ \\[0.1cm]
    $\begin{array}{|l|ccc|}
        \hline
        {}_{\sigma_i}\!\! \left\backslash \!{}^{\sigma_j} \!\! \right. & 1 & 2 & 3 \\
        \hline
        1 & E_1 & E_3 & E_2 \\
        2 & E_3 & E_4 & E_3 \\
        3 & E_2 & E_3 & E_1 \\
        \hline
    \end{array}$
\end{center}
}
\parbox{2.6cm}{
\begin{center}
    $h(\sigma_i,\sigma_j;3)$ \\[0.1cm]
    $\begin{array}{|l|ccc|}
        \hline
        {}_{\sigma_i}\!\! \left\backslash \!{}^{\sigma_j} \!\! \right. & 1 & 2 & 3 \\
        \hline
        1 & E_1 & E_2 & E_3 \\
        2 & E_2 & E_1 & E_3 \\
        3 & E_3 & E_3 & E_4 \\
        \hline
    \end{array}$
\end{center}
}
\end{center}
} \label{eqn-3-2}
\end{eqnarray}
Note that the $h(\sigma_i,\sigma_j;\alpha)$ are symmetric with respect 
to the diagonal, and
that they are related to each other through cyclic permutations,
i.e., from $\alpha = 1$ to $\alpha = 2$ and from $\alpha = 2$ to
$\alpha = 3$ dimer states have to be permuted cyclically, 
$1 \to 2 \to 3 \to 1$.

As  noted in~\cite{sarlah2005}, the four interaction energies do
not enter the phase diagram independently, but only two linear
combinations of these are relevant for the spectrum of
excitations. Introducing the thermal scale $k_B T$, a
two-dimensional phase diagram characterizes the complete
orientational ordering scenario of the dimer system. Here we
provide the full chain of arguments leading to this reduction of
control parameters. First it is convenient to sum over
the three
inequivalent lattice orientations separately, i.e., to split the
hamiltonian into ${\cal H} = \sum_\alpha {\cal H}_\alpha$ with
${\cal H}_\alpha = \sum_{\langle ij \rangle_\alpha}
h(\sigma_i,\sigma_j;\alpha)$. Then the corresponding hamiltonian
for each direction can be rewritten in a ``spin nomenclature'' as
a generalized spin-1-Ising or generalized Potts model. Here we
present the derivation for $\alpha = 1$, the results for the
remaining directions follow by cyclic permutations. The local
energy functional can be represented as a sum of four simpler
functionals
\begin{eqnarray}
    h(\sigma_i,\sigma_j;1) = \sum_{r=1}^4 h_r(\sigma_i,\sigma_j;1) ,
\label{eqn-3-3}
\end{eqnarray}
where the $h_r(\sigma_i,\sigma_j;1) \equiv h_r$ are given in terms
of tables
\begin{eqnarray}
\parbox{8.5cm}{
\parbox{0.75cm}{$h_1:$}
\parbox{3.5cm}{
    $\begin{array}{|l|ccc|}
        \hline
        {}_{\sigma_i}\!\! \left\backslash \!{}^{\sigma_j} \!\! \right. & 1 & 2 & 3 \\
        \hline
        1 & -K & 0 & 0 \\
        2 & 0 & -K & 0 \\
        3 & 0 & 0 & -K \\
        \hline
    \end{array}$
}
\parbox{0.75cm}{$h_2:$}
\parbox{3.0cm}{
    $\begin{array}{|l|ccc|}
        \hline
        {}_{\sigma_i}\!\! \left\backslash \!{}^{\sigma_j} \!\! \right. & 1 & 2 & 3 \\
        \hline
        1 & 2L & L & L \\
        2 & L & 0 & 0 \\
        3 & L & 0 & 0 \\
        \hline
    \end{array}$
}

\vspace{0.5cm}
\parbox{0.75cm}{$h_3:$}
\parbox{3.cm}{
    $\begin{array}{|l|ccc|}
        \hline
        {}_{\sigma_i}\!\! \left\backslash \!{}^{\sigma_j} \!\! \right. & 1 & 2 & 3 \\
        \hline
        1 & -M & 0 & 0 \\
        2 & 0 & 0 & 0 \\
        3 & 0 & 0 & 0 \\
        \hline
    \end{array}$
}
\parbox{0.75cm}{$h_4:$}
\parbox{3cm}{
    $\begin{array}{|l|ccc|}
        \hline
        {}_{\sigma_i}\!\! \left\backslash \!{}^{\sigma_j} \!\! \right. & 1 & 2 & 3 \\
        \hline
        1 & E_2 & E_2 & E_2 \\
        2 & E_2 & E_2 & E_2 \\
        3 & E_2 & E_2 & E_2 \\
        \hline
    \end{array}$
} } \label{eqn-3-4}
\end{eqnarray}
and each contribution is parameterized in terms of a single energy
scale,
\begin{eqnarray}
    K &=& -(E_1-E_2) , \nonumber\\
    L &=& E_3-E_2 , \label{eqn-3-6}\\
    M &=& -(E_4-E_1-2E_3+2E_2) \, . \nonumber
\end{eqnarray}
Due to the generic ordering of the interaction energies of 
colloidal dimers, the
parameters $K$ and $L$ are expected to be positive, while  $M$
may attain both signs in general. 
The functionals allow for a direct interpretation: 
First, $h_1$ favors dimers to
be in the same state, second, 
$h_2$ gives an extra contribution if a dimer is aligned
with the bond vector connecting the dimers. Third, $h_3$ 
accounts for the high symmetry configuration, viz. both dimers aligned with 
the bond vector. Fourth, $h_4$ is independent of the 
configuration and corresponds to global total shift of the energy. 
The functionals $h_r(\sigma_i,\sigma_j;\alpha)$ are easily reexpressed in terms
of Kronecker symbols, and one obtains for the local energy functional 
\begin{eqnarray}\label{eqn-3-5}
    h(\sigma_i,\sigma_j;\alpha)
        &=& -K \delta_{\sigma_i,\sigma_j}
    +L (\delta_{\sigma_i,\alpha}  + \delta_{\sigma_j,\alpha})\nonumber \\
& &  -M \delta_{\sigma_i,\alpha} \delta_{\sigma_j,\alpha}
        + E_2 \, .
\end{eqnarray}
Collecting results the total hamiltonian reads
\begin{eqnarray}
    {\cal H} &=& \sum_{\alpha = 1}^3 \sum_{\langle ij \rangle_\alpha}
        [- K \delta_{\sigma_i,\sigma_j}
        - M \delta_{\sigma_i,\alpha} \delta_{\sigma_j,\alpha}] \nonumber\\
    && + \sum_{\alpha = 1}^3 \sum_{\langle ij \rangle_\alpha}
        [L (\delta_{\sigma_i,\alpha} + \delta_{\sigma_j,\alpha}) + E_2] .
\label{eqn-3-7}
\end{eqnarray}
The second line of the previous relation evaluates to a constant
once the sum over all pairs and directions is performed,
$\sum_{\alpha} \sum_{\langle ij \rangle_\alpha}
[L (\delta_{\sigma_i,\alpha} + \delta_{\sigma_j,\alpha}) + E_2]
= N (E_2 + 2 E_3)$. The phase behavior is unaffected by this global
shift of energy, thus, we discard the constant, and after rearranging
terms, one arrives at the final expression for colloidal dimer
hamiltonian
\begin{equation}
    {\cal H} = - K \sum_{\langle ij \rangle}
        \delta_{\sigma_i,\sigma_j}
        - M \sum_{\alpha = 1}^3 \sum_{\langle ij \rangle_\alpha}
        \delta_{\sigma_i,\alpha} \delta_{\sigma_j,\alpha} .
\label{eqn-3-8}
\end{equation}
The excitation spectrum of orientational flips is governed solely
by the two energy scales $K,M$ to be compared to the thermal one
$k_B T$, and the phase diagram reduces to a two-dimensional plane. 
The $K$-term accounts for the energy gained if two neighboring dimers 
are in the same orientational state, see Fig.~\ref{fig-3-1}(b), 
$E_4$ or $E_1$. The $M$-term distinguishes 
the high symmetry state of alignment of the dimers with 
the lattice bond vector, Fig.~\ref{fig-3-1}(b), $E_4$. This coupling between
spin configuration and orientation of the bond vector has a close analogue in
 the magnetic dipole-dipole interaction of localized spins on a lattice.

\vspace{0.5cm} 
The spin hamiltonian of Eq. (\ref{eqn-3-8}) can be
interpreted as a generalization of a $q$-state {\em Potts model},
the natural extension of the Ising model. In these systems, each
``spin'' $i$ can be in one of the $\sigma_i = 1,..,q$\, equivalent
states. The only energetic contribution occurs if two neighboring
spins are in the same state, i.e.,
\begin{equation}
    {\cal H} = - J \sum_{\langle ij \rangle}
        \delta_{\sigma_i,\sigma_j} .
\label{eqn-3-9}
\end{equation}
It is obvious that the colloidal dimer hamiltonian reduces to a
three-states Potts model if $M=0$.

There have been intensive studies of the critical properties of
ferromagnetic, $J>0$, as well as anti-ferromagnetic, $J<0$, Potts
models for various lattice geometries in two and three dimensions.
In particular, a rigorous solution of the ferromagnetic Potts model
in two dimensions is available~\cite{baxter1973,wu1982} for a
square, triangular, and honeycomb lattice for $q = 2$ (Ising) and
$q \geq 4$. The melting of the ferromagnetic phase to the
paramagnetic one occurs via a continuous phase transition for the
case of $q = 2,4$ and via a discontinuous one for $q > 4$. For
both cases the exact value of the critical temperature for the
triangular lattice is given by~\cite{baxter1989,wu1982}
\begin{eqnarray}
    \exp \frac{J}{k_B T_c} &=&
        2 \cos\left({\frac{2}{3}
            \cos^{-1}\left(\frac{\sqrt{q}}{2}\right)}\right)
        \, ; \  q = 2,4 \label{eqn-3-10}\\
    \exp \frac{J}{k_B T_c} &=&
        2 \cosh\left(\frac{2}{3}
            \cosh^{-1}\left(\frac{\sqrt{q}}{ 2}\right)\right)
        \, ; \  q \geq 4 \nonumber \, .
\end{eqnarray}
Unfortunately, for the interesting case of $q = 3$ neither the
method of transfer matrices (for the Ising model) nor the circle
theorem~\cite{baxter1989} for the vertex model (for  $q \geq 4$ Potts models) is
applicable. Nevertheless, since the critical point in
Eq.~(\ref{eqn-3-10}) agrees with the exact Ising result, it is
expected that it holds also for $q = 3$. There is also  strong
numerical evidence by Monte Carlo simulations \cite{wu1982,binder1981} that the critical
temperature is still given by Eq.~(\ref{eqn-3-10}). Furthermore,
Monte Carlo simulations as well as renormalization group analysis
corroborate a continuous transition scenario~\cite{wu1982} for $q=3$.

It is interesting to note that a standard mean-field analysis
gives a qualitatively wrong result as far as the nature of the
transition is concerned, i.e., a discontinuous scenario is
predicted for $q \geq 3$. The value of the critical temperature is
overestimated,
\begin{equation}
    k_B T_\text{MF} = {3 J \over 2 \ln2} ,
\label{eqn-3-11}
\end{equation}
with respect to the tentatively exact value
$k_B T_c = J / \ln[2 \cos (\pi/9)]$ by a factor of $1.3654$ similar
to the case of the 2D Ising model. Since we recover the three-state
Potts model as a special case of the colloidal dimer hamiltonian,
Eq.~(\ref{eqn-3-8}), one expects that the mean-field analysis fails to
predict the correct nature of the transitions also in the case of
the colloidal dimer hamiltonian.

For negative exchange coupling $J$ with a low-temperature
anti-ferromagnetic state there are no exact solutions known. For
triangular lattices renormalization group (RG)
studies~\cite{schick1977}, series analysis~\cite{enting1982}, and
Monte Carlo simulations~\cite{saito1982} give for the critical
point a numerical estimate $k_B T_c \approx J/\ln{(0.206)}$. On
the other hand, there is no general consensus on the nature of the
transition. Extrapolated series expansions and  Monte Carlo
simulations find the transition  to be discontinuous, whereas
renormalization group analysis yields a continuous phase change~\cite{wu1982}.

\subsection{Mean-field description}\label{mean-field}
A rigorous solution of the colloidal dimer problem is certainly
desirable, however, it would imply an analytic solution of the
three-state Potts model as a special case, a problem that remains
unsolved despite considerable efforts~\cite{wu1982,baxter1989}.
Approximate methods are required to gain further insight in the
the phase behavior. Mean-field theory has proven a
powerful tool that allows to identify the global topology of the
phase diagram, in particular, the coexistence of phases of
different broken symmetries, as well as to provide approximate
numbers for the location of phase boundaries. The results of such
a mean-field analysis also provide a useful reference for a
numerical simulation by Monte Carlo methods which will be
presented in Sec.~\ref{MonteCarlo}.

We use the framework of {\em variational mean-field theory} and
determine the optimal distribution of dimer states within a
restricted class of density matrices~\cite{chaikin1995}. The
normalized equilibrium density matrix $\rho  = Z^{-1}\exp(-\beta
{\cal H})$, with inverse temperature $\beta = 1/k_B T$, is
difficult to obtain since the partition sum $Z = \, {\rm Tr} \,[\exp(-\beta
{\cal H})]$ usually cannot be evaluated exactly. To overcome this
problem it is favorable to approximate $\rho$ by a product of
single-site density matrices $\rho_i$,
\begin{equation}
    \rho (\{\sigma_i\}) = \prod_i \rho_i (\sigma_i) ,
\label{eqn-4-1}
\end{equation}
i.e., spins at different sites are taken as uncorrelated. The
quantities $\rho_i (\sigma_i)$ represent the probability to find
the dimer at site $i$ in  state $\sigma_i$. Normalization implies
in particular $\sum_{\sigma_i} \rho_i(\sigma_i) = 1$. The free
energy corresponding to a system represented by a density matrix
as in  Eq.~(\ref{eqn-4-1}) results from the usual balance energy
vs. entropy ${\cal F}_\rho = {\cal E}_\rho - T {\cal S}_\rho$,
where the mean energy ${\cal E}_\rho$ and the entropy ${\cal
S}_\rho$ are evaluated by the rules of statistical mechanics
\begin{equation}
    {\cal E}_\rho = \langle {\cal H} \rangle_\rho \, , \qquad
        {\cal S}_\rho = -k_B  \langle \ln \rho \rangle_\rho \, ,
\label{eqn-4-2}
\end{equation}
where $\langle \dots \rangle_\rho$ denotes averaging with respect
to the approximate distribution of Eq.~(\ref{eqn-4-1}). One can
show in general that Bogolyubov's inequality~\cite{chaikin1995} 
sets a lower bound
to ${\cal F}_\rho$,
\begin{equation}
    {\cal F}_\rho = \, {\rm Tr} \,[\rho {\cal H}] + k_B T \, {\rm Tr} \,[\rho \ln\rho]
        \geq {\cal F} \, .
\label{eqn-4-3}
\end{equation}
Here ${\cal F}$ denotes the exact thermodynamic free energy. The idea of
the variational mean-field method is to find the single-particle
$\rho_i(\sigma_i)$ that yields the variational mean-field free
energy ${\cal F}_\rho$ closest to the thermodynamic one, viz.
${\cal F}$.

For the colloidal dimer hamiltonian the variational mean-field
free energy reads
\begin{eqnarray}
 {\cal F}_\rho & = &- K \sum_{\langle ij \rangle} \sum_\sigma
   \rho_i (\sigma) \rho_j (\sigma)
  - M \sum_\alpha \sum_{\langle ij \rangle_\alpha}
  \rho_i (\alpha) \rho_j (\alpha) \nonumber \\
 & & + k_B T \sum_i \sum_\sigma \rho_i (\sigma) \ln{[\rho_i (\sigma)]} \, .
\label{eqn-4-4}
\end{eqnarray}
Note that in the $M$-term the spin orientation and the lattice bond are 
aligned, $\alpha= \sigma$. 
This expression for the free energy is still subject 
to the minimization procedure with
respect to appropriate sets of single-site density matrices.

The symmetries of the dimer states in the broken symmetry phase
have to be reflected in the ansatz for the $\rho_i(\sigma_i)$. In
general, one introduces a number of sublattices each comprising a
set of equivalent lattice sites with respect to orientational
order.

In the simplest case, all lattice sites are equivalent and the
only phase transition allowed separates a {\em paramagnetic} phase from
a {\em ferromagnetic} one. The suitable density matrix is
therefore site-independent,
\begin{equation}
    \rho_i (\sigma_i) \equiv \rho (\sigma_i) \mapsto \rho_{\sigma_i} .
\label{eqn-4-5}
\end{equation}
Upon substituting this ansatz in the mean-field free energy,
Eq.~(\ref{eqn-4-4}), yields the free energy density, i.e., free
energy per lattice site, $f = {\cal F}_\rho / N$,
\begin{equation}
    f = - (3 K + M) \sum_\sigma \rho_\sigma^2
        + k_B T \sum_\sigma \rho_\sigma \ln \rho_\sigma .
\label{eqn-4-6}
\end{equation}
Let us note that within the single-site mean-field description,
the parameters $K$ and $M$ appear only in the combination $3K +
M$. Applying the same procedure for the three-state Potts model
gives an identical expression for the free energy density provided
$3 K + M$ is substituted by $J$.

A ferromagnetic phase (FM) is characterized by a broken symmetry,
i.e., a particular dimer state is preferred with respect to the
remaining. The single dimer excitations with respect to the ferromagnetic
ground state are degenerate, i.e., the two possible minority
orientations are equally probable. By the symmetry of the FM, this
property is preserved in the whole phase region. Selecting the
majority component as the $\sigma = 1$ state, we parametrize 
$\rho_1=(1+2m)/3$ by a generalized 
magnetization $m$. The minority probabilities are determined by the 
normalization 
$\sum_\sigma \rho_\sigma = 1$ to 
$\rho_2=\rho_3=(1-m)/3$. The magnetization is restricted to 
$-1/2 \leq m \leq 1$, and 
the fully aligned ferromagnetic state corresponds 
to $m=1$, whereas the paramagnetic one is given by $m=0$. 
A negative magnetization would indicate that the distinguished spin state is
less probable than the remaining two. Hence there is an inherent  
asymmetry $m \mapsto -m$ in the mean-field description, and as we shall 
see later, all transitions
within this mean-field approach are (falsely) predicted to be discontinuous,
similar to the case of a pure Potts model.

\begin{figure}[htbp]
\includegraphics[width=1.0\linewidth]{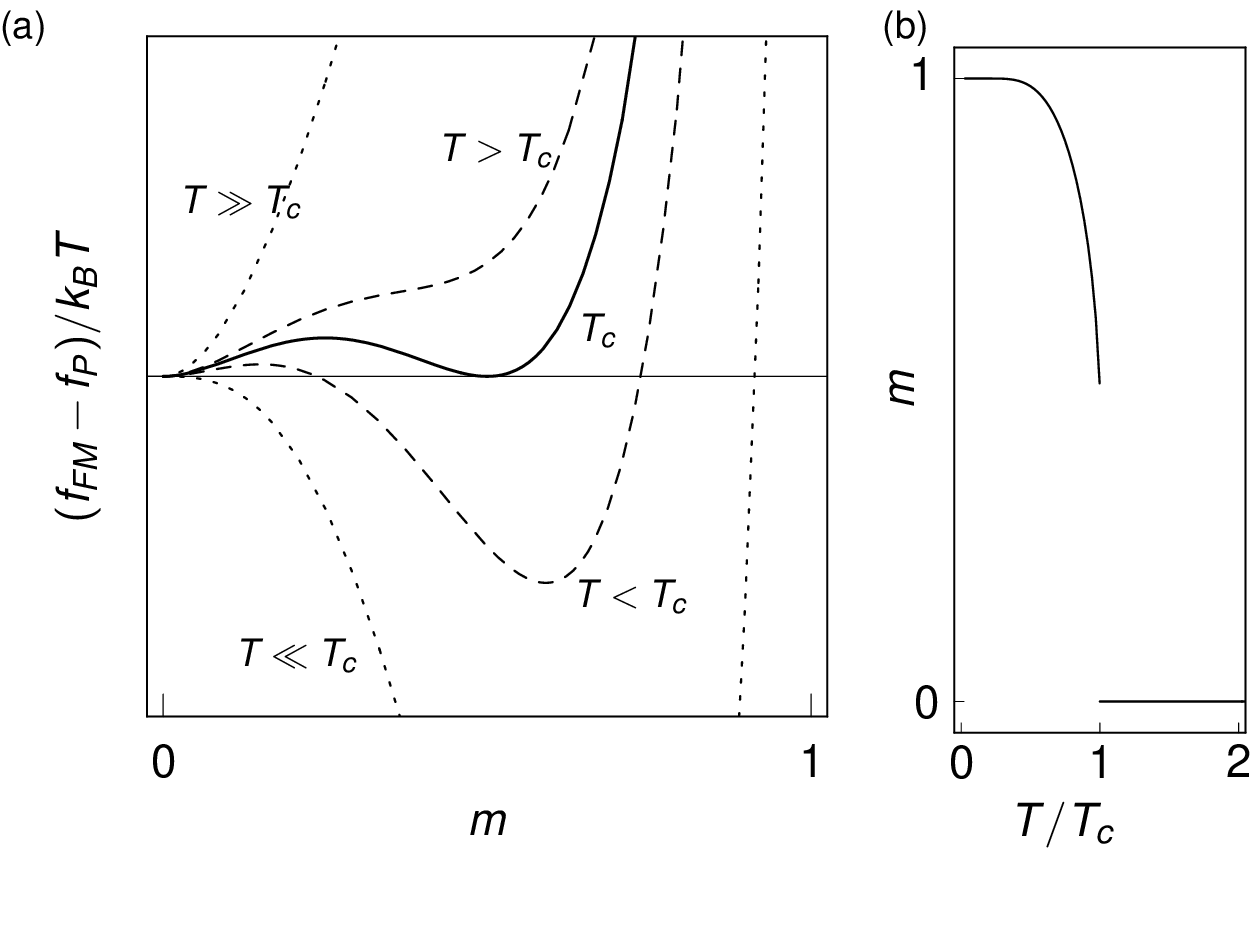}
\vspace{-1.25cm}
\caption{a) Free energy as a function of the magnetization for 
various temperatures above, below, and at the critical point. 
Local minima correspond to metastable phases. b) Mean-field 
temperature dependence of the magnetization for the ferromagnetic 
phase of the colloidal dimer hamiltonian. At the critical 
temperature, $T/T_c=1$, the order parameter exhibits a jump.}
\label{fig-4-1}
\end{figure}

Substitution into Eq.~(\ref{eqn-4-6})
yields the excess free energy of the ferromagnetic phase with
respect to paramagnetic one,
\begin{eqnarray}
    {f_{\rm FM}-f_{\rm P} \over k_B T} &=&
        - {2 \epsilon \over 3} m^2
        + {1+2m \over 3} \ln{(1+2m)} \nonumber\\
        && + 2 {1-m \over 3} \ln{(1-m)} \, ,
\label{eqn-4-9}
\end{eqnarray}
which depends on the effective parameter $\epsilon = \beta(3 K +
M)$. The corresponding graph is shown in Fig.~\ref{fig-4-1}~(a). 
One infers the typical characteristics of a first order phase
transition, e.g., a discontinuous change of the order parameter
and metastable states.

The equation of state for the magnetization is obtained by
minimizing $f_{\rm FM}$ with respect to $m$,
\begin{eqnarray}
    2 \epsilon m = \ln{1+2m \over 1-m} \, .
\label{eqn-4-8}
\end{eqnarray}
One infers that $m=0$, corresponding to the paramagnetic phase, is
always a solution to this equation. Depending on the value of the
effective parameter $\epsilon$ non-trivial solutions may occur
that exhibit lower mean-field free energies than the paramagnetic
one, $f_{\rm P} = - k_B T (\epsilon/3 - \ln 3)$. For infinite
temperatures ($\epsilon = 0$) the solution to this equation
reduces to the paramagnet ($m=0)$, whereas at zero temperature
($\epsilon \to \infty$) full alignment of the ferromagnet ($m=1$)
occurs. The complete temperature dependence of the magnetization
is depicted in Fig.~\ref{fig-4-1}~(b).

\begin{figure}[htbp]
\includegraphics[width=1.0\linewidth]{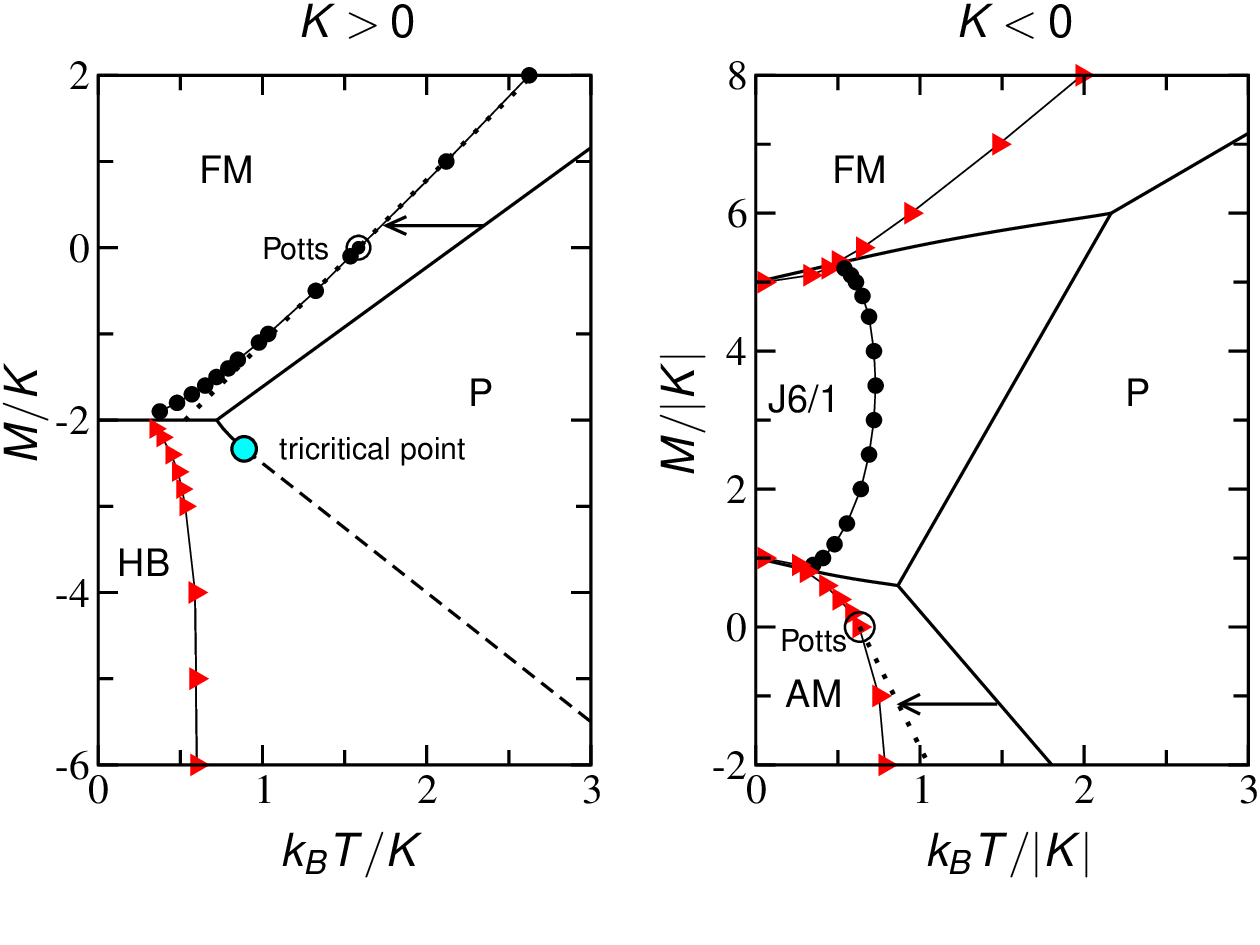}
\vspace{-1cm}
\caption{Phase diagram for colloidal dimers. Solid lines represent 
first order transitions derived from the MF description and the 
dashed line after the tricritical point represents a line of 
second order transitions. Dots and triangles denote the respective 
continuous and discontinuous transition points as obtained by MC 
simulations; the two encircled symbols indicate the pure Potts 
transitions. Dotted lines represent the extrapolation of the 
critical point of the Potts model, see text.}
\label{fig-4-5}
\end{figure}

The thermodynamic transition from a paramagnetic phase occurs at
the temperature $T_c$ when the non-trivial minimum of the free
energy $f_{\rm FM}$ equals the one of the paramagnetic state
$f_{P}$. With the help of the equation of state,
Eq.~(\ref{eqn-4-8}), one finds for the critical effective
parameter $\epsilon_c = 2 \ln 2$, which implies for the critical
temperature
\begin{eqnarray}
   {\rm FM-P}:\quad  k_B T_c = {3K+M \over 2 \ln 2 } \, .
\label{eqn-4-10}
\end{eqnarray}
Representing the phase diagram in terms of the dimensionless
parameters $ k_B T/|K|$ and $ M/|K|$, the FM--P phase boundary
yields a straight line, see Fig.~\ref{fig-4-5}.

The magnetization at the critical point is finite and attains the
value $m_c = 1/2$.
The jump in the derivative of free energy density with respect to
temperature at the phase transition implies a latent heat per
dimer which evaluates to $q_l/k_B T_c = \ln(2)/3$. The heat 
capacity jumps at $T_c$ by
$\Delta c_N = (8k_B/3) [\ln^2(2) / (3-4 \ln(2))]$. Note that 
these quantities are universal in the sense
that they do not depend on the parameters $K$ and $M$, i.e, they
are constant along the whole phase boundary. The mean-field
analysis appears to be qualitatively correct for $K<0$, whereas for
$K>0$ the simulation results
 suggest that
fluctuations render the phase transition to a continuous one, as
will be discussed in detail in Sec.~\ref{MonteCarlo}. 

\begin{figure}[htbp]
\includegraphics[width=1.0\linewidth]{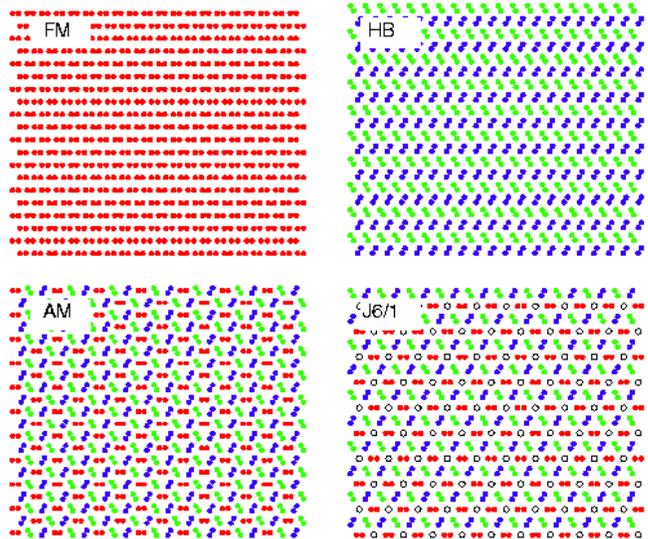}
\vspace{-1.75cm}
\caption{Ground state ordered structures of 2D colloidal dimers: 
(a) ferromagnetic (FM), (b) herring bone (HB), (c) Potts 
anti-ferromagnetic (AM), and (d) Japanese 6 in 1 (J6/1), structure. 
In the latter, the energy of the system does not depend on the 
orientation of dimers on sites denoted by circles.}
\label{fig-4-4}
\end{figure}

The ferromagnetic phase represents only the simplest broken
symmetry. Spatially varying order parameters give rise to complex
phase behavior, in particular, the sites fall into different
classes of sublattices. Comparing the exact ground state energies
of different ordered  structures, one can infer the possible
broken symmetry phase at low temperatures. In total, we have found
four ordered phases of different symmetry, see Fig.~\ref{fig-4-4},
that are thermodynamically stable in different regions of the
phase diagram, depicted in Fig.~\ref{fig-4-5}. In addition to the
ferromagnetic one, there appears an anti-ferromagnetic phase (AM)
stable for $K<0$ and not too large $M$. Here, in the fully ordered
state, neighboring dimers are in different orientational states.
On the triangular lattice this implies that each dimer is
surrounded by three neighbors of a configuration rotated by
$60^\circ$ and by three neighbors of a configuration rotated by
$-60^\circ$ with respect to the configuration of the central
dimer. There are three triangular sublattices with a lattice
constant of $a\sqrt{3}$ and the sublattices are equivalent in the
sense that each of them can be obtained  by a spatial translation
with a simultaneous rotation of the dimers of any other one. For
positive $K$ and sufficiently negative $M$ a herring bone (HB)
structure represents the state of lowest free energy, see
Fig.~\ref{fig-4-4}. Statistically equivalent dimers are arranged
in two sets of rows, the majority orientational states
corresponding to the ones not parallel to the orientation of the
rows. Finally, we have found that intervening in between the FM and
AM for $K<0$ there appears a phase of four triangular sublattices.
Three of them are equivalent in the sense mentioned above, the
fourth one exhibiting no preferred orientation. We refer to it as
Japanese 6 in 1 phase (J6/1), due to its resemblance to a weaving
pattern for chain mailles worn by samurais in the $14^{\small\text{th}}$
century~\cite{asa-no-ha-gusari}.

To describe these complex phases generalized ansatzes have to be
used for the density matrix. Then the site dependence of the
components of $\rho_i (\sigma_i)$ is encoded in the sublattice
$\Gamma$ the site $i$ belongs to, i.e.,
\begin{equation}
    \rho_i (\sigma_i) = \rho_{\sigma_i}^\Gamma ; \quad
        i \in~{\rm sublattice}~\Gamma \, .
\label{eqn-4-13}
\end{equation}
Some of the density matrices $\rho_{\sigma}^\Gamma$ are related by
permutations of dimer states. Explicitly, for the AM phase as well
as for the HB the sublattices are equivalent, so that only one
density matrix characterizes the order of the system. For the J6/1
phase the order in the three equivalent sublattices is determined
by one non-trivial density matrix, whereas the fourth one
attributes equal probabilities to all dimer orientations. Let us
mention an additional symmetry that occurs for the AM and J6/1
phase. Here, the two minority orientations are equivalent as was
the case in the FM phase. The corresponding density matrix is
parameterized by a single number, a generalized magnetization $m$
defined on a sublattice. The situation is different for the HB
phase. An excitation of a dimer to the majority orientation of the
neighboring row is in general different from aligning the dimer
with the orientation of the row. Correspondingly, the density
matrix in the HB phase requires two parameters.

The three sublattices for the {\em anti-ferromagnetic} phase are inferred 
from the ground state represented in Fig.~\ref{fig-4-4}.
The single-site density matrices  corresponding to the sublattices $A,B,C$, see
Eq.~(\ref{eqn-4-13}), are parameterized by
\begin{eqnarray}
    \rho^A_1 &=& \rho^B_2 = \rho^C_3 = (1+2m)/3 \, , \nonumber \\
    \rho^A_{2,3} &=& \rho^B_{1,3} = \rho^C_{1,2}  = (1-m)/3 \, .
\label{eqn-4-14}
\end{eqnarray}
The excess free energy density of the anti-ferromagnet is then
obtained by substituting this ansatz into Eq.~(\ref{eqn-4-4}),
\begin{eqnarray}
    {f_{\rm AM} -f_{\rm P}\over k_B T} &=&
        - {2 \mu \over 3} m^2
        + {1+2m \over 3} \ln{(1+2m)} \nonumber\\
    && + 2 {1-m \over 3} \ln{(1-m)} \, ,
\label{eqn-4-15}
\end{eqnarray}
where the sole parameter entering this expression is given by $\mu
= -\beta(3K+M)/2$. The shape of the free energy density $f_{\rm
AM}$ as a function of temperature is identical  to the one of the
ferromagnetic phase $f_{\rm FM}$, see Fig.~\ref{fig-4-1}. The
equation of state for the sublattice magnetization $m$ is thus
obtained by replacing $\epsilon$ by $\mu$ in Eq.~(\ref{eqn-4-8}),
i.e., $ 2 \mu m = \ln[(1+2m)/( 1-m)]$.
The same substitution for the critical temperature,
Eq.~(\ref{eqn-4-10}), yields
\begin{eqnarray}
    {\rm AM-P}:\quad k_B T_c = - {3 K + M \over 4 \ln 2} \, .
\label{eqn-4-17}
\end{eqnarray}
The AM--P phase boundary is represented by a straight line in the
$T/|K|$---$M/|K|$ phase diagram, Fig~\ref{fig-4-5}. The latent
heat and the excess of the heat capacity are again the same as for
the FM phase provided one substitutes $\epsilon$ with $\mu$.

For the {\em Japanese 6 in 1} phase the appropriate ansatz for the
three equivalent sublattices ($A,B,C$) is identical to the
anti-ferromagnetic case, Eq.~(\ref{eqn-4-14}), whereas the fourth
($O$) is degenerate
\begin{eqnarray}
    \rho^O_{1,2,3} = 1/3 \, .
\label{eqn-4-18}
\end{eqnarray}
The excess free energy density for the J6/1 symmetry is readily
obtained
\begin{eqnarray}
    {f_{\rm J6/1}-f_{\rm P} \over k_B T} &=& {3 \over 4}
        \left[ -{2 \nu \over 3} m^2
        + {1+2m \over 3} \ln{(1+2m)}
        \right.\nonumber \\
    && \left. + 2 {1-m \over 3}\ln{(1-m)} \right]\, .
\label{eqn-4-19}
\end{eqnarray}
Here $\nu = -\beta (3K-M)/3$ denotes the sole effective parameter.
Up to the multiplicative constant $3/4$ the excess free energy
density of the J6/1 phase is identical to the excess free energy
densities of the FM and AM phases discussed above, provided $\nu$
is substituted by $\epsilon$ or $\mu$, respectively. The prefactor
$3/4$ arises since only three of the four sublattices exhibit a
broken symmetry. The equation of state obtained as the derivative
of the free energy density with respect to the sublattice
magnetization $m$ is again formally identical to the cases
discussed above,
    $2 \nu m = \ln[(1+2m) / (1-m)] $.
 The critical
temperature for the J6/1--P phase transition is readily obtained
via the substitution rule, yielding $\nu_c = 2 \ln 2$ and
\begin{eqnarray}
    {\rm J6/1-P}:\quad k_B T_c = - {3K-M \over 6 \ln 2} \, .
\label{eqn-4-21}
\end{eqnarray}
The J6/1--P phase boundary is indicated in  Fig.~\ref{fig-4-5}.
Since only three of the four sublattices exhibit a broken
symmetry, the latent heat as well as the jump of the heat capacity
at the phase transition are $3/4$ of the universal values of the
FM--P or AM--P transitions.

The {\em herring bone} structure (HB) requires two parameters, one
for the majority magnetization $m$ and a second 'biaxiality'
parameter, $n$, for the asymmetry of the minority orientations.
 The need for the biaxiality
parameter becomes clear if one considers excitations of HB ground
state. A single dimer flip to the orientation of the neighboring
row costs an energy of $\Delta E =-2K-2M $, whereas aligning it
with the row induces an energy change of only $\Delta E = 2K$.
(Note that the single dimer excitation energies become degenerate
for $M = -2K$.) The low temperature excitations of the HB
structure are represented by a dilute gas of defects of the latter
kind. Considering the sign of the single-dimer excitation energies
an estimate for the stability of the HB structure at low
temperatures is possible, i.e., for positive $K$ and sufficiently
large negative $M$.

\begin{figure}[htbp]
\includegraphics[width=1.0\linewidth]{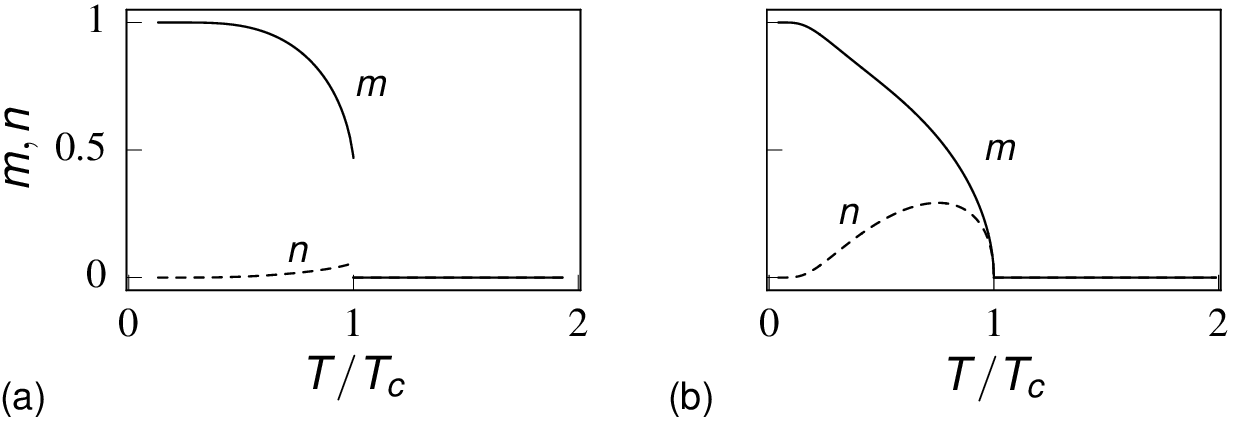}
\caption{Order parameters $m$ and $n$ as a function of temperature
for a path of constant $M/K$. (a) A discontinuous phase transition 
characterized by a jump of $m$ and $n$ for $M/K=-2.1$ and (b) a 
continuous phase transition for $M/K=-5$.}
\label{fig-4-3}
\end{figure}

For the parameterization of the sublattice single-site density
matrices we choose
\begin{eqnarray}
    \rho^A_2 &=& \rho^B_3 = (1+2m)/3 \, ,\nonumber \\
    \rho^A_{1,3} &=& \rho^B_{1,2} = (1-m \pm n)/3 \, .
\label{eqn-4-22}
\end{eqnarray}
Then the free energy density corresponding to the HB symmetry
reads
\begin{eqnarray}
    {f_{\rm HB}-f_{\rm P} \over k_B T} &=&  {M \over 3 k_B T} m^2
        - {4K+M \over 9 k_B T} n^2 \nonumber \\
    && + {2 (2K+M) \over 3 k_B T} m n
            + {1+2m \over 3} \ln{(1+2m)} \nonumber \\
        && + {1-m+n \over 3} \ln{(1-m+n)} \nonumber \\
        && + {1-m-n \over 3} \ln{(1-m-n)} \, .
\label{eqn-4-23}
\end{eqnarray}
The excess free energy is subject to a simultaneous  minimization
with respect to both order parameters $m,n$. In general, one
cannot find an analytic solution of the equations of state
\begin{eqnarray}
& &  k_B T \ln{1-m+n \over 1-m-n}  =  - 2 (2K + M) m + {2 \over 3}
(4K+M) n \, , \nonumber\\
   & & k_B T \ln{(1+2m)^2 \over (1-m)^2-n^2} =  - 2 M m - 2 (2K+M) n
     \, .
\label{eqn-4-24}
\end{eqnarray}
The generic behavior close to the HB--P phase transition is
illustrated in Fig.~\ref{fig-4-3} for a path of constant $M/|K|$
close to the HB--FM line and one starting deep in the HB phase.

For the case of a continuous transition the transition temperature
can be obtained analytically by expanding the excess free energy
density in powers of the small quantities $m$ and $n$, which up to
quadratic order reads
\begin{eqnarray}
\frac{f_{\rm HB}-f_{\rm P}}{k_B T} &=&
  \left(\frac{M}{3 k_B T} + 1 \right) m^2
    + \frac{4 K + 2 M}{3k_B T} mn \nonumber \\
    & & + \left(1 - \frac{4 K + M}{3k_B T} \right) \frac{n^2}{3}
         \, .
\label{eqn-4-25'}
\end{eqnarray}
The quadratic form ceases to be positive definite and the
paramagnetic phase becomes unstable via a continuous transition at
the critical temperature
\begin{eqnarray}
    {\rm HB-P},\ M \leq -\frac{7K}{3} :\quad k_B T_c = - {2 (K+M) \over 3} \, ,
\label{eqn-4-26}
\end{eqnarray}
and the critical direction is along the diagonal $(m,n) \sim
(1,1)$, i.e the order parameters become equal close to the
transition, see Fig. \ref{fig-4-3}. In the regime $-2 K \leq M
\leq -7K/3$ the melting of the HB structure towards the
paramagnetic phase is discontinuous. The {\em tricritical} point,
 where the order of the phase
transition changes can by determined again by local
considerations. A suitable non-linear transformation of variables
$(m,n)$, shows that the origin becomes unstable with respect to
higher order terms along the critical direction at $M = -7 K/3, \,
k_B T_{3c} = 8 K/9$.

For $M=-2K$ the excitation energies of the isolated single-dimer
flips in the HB structure become degenerate. This is reflected by
the fact that the excess free energy, Eq.~(\ref{eqn-4-23}),
exhibits the additional symmetry $n \to - n$ along this line. At
zero temperature the HB phase is perfectly ordered $(m=1,n=0)$,
i.e in a $n$-symmetric phase. Increasing the temperature does not
give rise to a continuous or discontinuous $n$-symmetry breaking,
so $n\equiv 0$ holds along the $M=-2K$ line for all temperatures.
With this observation in mind, one checks that the mean-field free
energy density of the HB phase for $M = -2 K$ equals the one of
the FM phase for all temperatures. Thus, the line $M = -2 K$ is a
coexistence line of the two neighboring phases. The special point
for the discontinuous transition to the P phase, $ k_B T_c = K/(2
\ln{2})$, constitutes a triple point where FM, HB, and P phases
coexist.

The remaining {\em coexistence lines} between different ordered
phases, i.e., for $K < 0$ between FM and J6/1 and between AM and
J6/1, can only be determined numerically and are included in the
phase diagram. Here we present in short approximate analytical 
solutions in terms of asymptotic low-temperature expansions.

At the coexistence line the free energy densities of the two
coexisting phases are equal. In particular, for the FM--J6/1 line
$f_{\rm FM}(m_{\rm FM},T) = f_{\rm J6/1}(m_{\rm J6/1},T)$, where
$f_{\rm FM}$ and $f_{\rm J6/1}$ are defined in
Eqs.~(\ref{eqn-4-9}) and (\ref{eqn-4-19}) and the $m$'s are
solutions of the corresponding equations of state. For $T \to 0$
($M/|K| \gtrsim 5$), $m_{\rm FM}, m_{\rm J6/1} \approx 1$ which
yields the coexistence line
\begin{eqnarray}
    {\rm FM-J6/1},\ M/|K| \gtrsim 5:\quad
        k_B T_c \approx {2 (5 K + M) \over \ln 3} \, .
\label{eqn-4-27}
\end{eqnarray}
Close to the FM--J6/1--P triple point, $ M = 6 |K|$, $ k_B T = 3
|K| /(2 \ln{2})$, the magnetizations approximately take their
critical values, $m_{\rm FM}, m_{\rm J6/1} \sim m_c = 1/2$, and
give rise to
\begin{eqnarray}
    {\rm FM-J6/1},\ M/|K| \lesssim 6:\quad
        k_B T_c = {3 (5 K + M) \over 2 \ln{2}} \, .
\label{eqn-4-28}
\end{eqnarray}
For the AM--J6/1 line an equivalent procedure with the same
approximate solutions for the magnetizations close to the ground
state or triple point results in
\begin{eqnarray}
    {\rm AM-J6/1}, M/|K| \lesssim 1:\quad
        k_B T_c \approx - {2 (K + M) \over \ln 3} \, , \\
    M/|K| \gtrsim {3 \over 5}:\quad
        k_B T_c \approx - {3 (K + M) \over 2 \ln 2} \, .
\label{eqn-4-29}
\end{eqnarray}
Analytical expressions for exact or approximate critical lines are
gathered in Tab.~\ref{tab-4-1} and the mean-field special lines,
i.e., tricritical and triple points, in Tab.~\ref{tab-4-2}.

\begin{table}
\caption{Phase boundaries obtained in mean-field theory. The approximate 
results for the J6/1--FM and J6/1--AM transition correspond to the asymptotic
low-temperature (high-temperature) solution.} \medskip
\begin{tabular}{|l|l|}
\hline 
transition & critical line\\ 
\hline 
$K > 0$ & \\ \hline
FM--P & $ k_B T_c = (3 K + M) / (2 \ln 2)$ \\ 
HB--P & $ k_B T_c = -(2/3) (K + M)$, $M \leq -7 K/3$ \\ 
FM--HB & $ M = -2 K$ \\ 
\hline 
$K < 0$ & \\ 
\hline 
FM--P & $k_B T_c = (3 K + M) / (2 \ln 2)$ \\ 
J6/1--P & $k_B T_c = (-3 K + M) / (6 \ln 2)$ \\ 
AM--P & $k_B T_c = - (3 K + M) / (4 \ln 2)$ \\
J6/1--FM & $k_B T_c \approx 3 (5 K + M) / (2 \ln 2)$,
    $M \lesssim 6 |K|$ \\
& $k_B T_c \approx 2 (5 K + M) / \ln 3$,
    $M \gtrsim 5 |K|$ \\
J6/1--AM & $k_B T_c \approx - 3 (K + M) / (2 \ln 2)$,
    $M \gtrsim 3|K|/5$ \\
& $k_B T_c \approx - 2 (K + M) / \ln 3$,
    $M \lesssim |K|$ \\
\hline
\end{tabular}
\label{tab-4-1}
\end{table}

\begin{table}
\caption{Mean-field special points.}
\begin{tabular}{|l||r@{,}l|}
\hline 
coexisting phases & [$k_B T/|K|$ & $M/|K|$] \\ 
\hline
\hline 
$K > 0$ & \multicolumn{2}{|c|}{} \\ 
\hline 
FM--HB--P &[$1/(2 \ln{2})$ & $-2\,$] \\ 
HB--P tricritical point & [8/9 &-7/3]\\ 
\hline 
$K < 0$ & \multicolumn{2}{|c|}{} \\ 
\hline
FM--J6/1--P & [$3/(2 \ln{2})$ & $\ 6\,$] \\ 
AM--J6/1--P & [$3/(5\ln{2})$ & $\, 3/5$] \\ 
\hline
\end{tabular}
\label{tab-4-2}
\end{table}

\subsection{Results of the Monte Carlo simulations}
\label{MonteCarlo}
In this subsection we exemplify the phase behavior of the
colloidal dimer system by numerical exact means, i.e., Monte Carlo
methods. In particular, we corroborate the existence of all the
phases discussed within the mean-field analysis and address the
nature of the phase boundaries separating the corresponding phases
of broken symmetry. The results obtained within mean-field thereby
play a crucial role to identify the relevant order parameters to
be monitored.

The standard Metropolis algorithm~\cite{metropolis1953} has been
employed to generate a sequence of configurations representative
for the canonical ensemble to some specified temperature. The
starting configuration for the lowest temperature has been chosen
as the perfectly ordered (exact) ground state corresponding to the
pair of energy scales $(K,M)$. The temperature has been increased
gradually starting from the representative configuration of the
previous parameter set. We have considered systems of different
sizes subject to periodic boundary conditions. Explicitly, the
number of rows $L$, each consisting of $L$ dimers, has been
varied, $L=12, 32, 52, 102$, to ensure that the results have
approached the infinite system size limit. Before actual data were
collected the system has been carefully equilibrated by performing
up to several thousand  Monte Carlo cycles. Each Monte Carlo
cycle consists of $N \ln N$ \footnote{Within Monte Carlo simulations 
the term {\em Monte Carlo cycle} denotes the number of attempted moves 
one should perform to assure that on average each spin is selected 
once. Although the problem of finding the appropriate number (known as 
{\em Coupon collector's problem}) is exactly solved, see 
e.g.~\cite{grimmett1995}, the factor $\ln N$ is often omitted in the 
physics literature. In simulations, this is then ``corrected'' by either 
analysing a huge numbers of configurations or by sampling only 
configurations being separated a certain number of cycles. In the 
vicinity of phase transitions with pronounced correlations this has to 
be done anyway.}
attempted dimer flips, where $N = L^2$
is the total number of dimers in the system. To ensure that the
configurations entering the data analysis are statistically
independent, we have considered configurations separated by up to
1000 Monte Carlo cycles in the vicinity of a phase transition.

To characterize the broken symmetry phases we have considered the
respective order parameters. For example, in the ferromagnetic
case the fluctuating magnetization reads
\begin{eqnarray}
    {\cal M} = {{\cal N}_\Sigma \over N}
        - {1 \over 2}
        {\sum_{\sigma \ne \Sigma} {\cal N}_\sigma \over N} \, .
\label{eqn-5-1}
\end{eqnarray}
Here, ${\cal N}_\sigma$ is the fluctuating number of dimers in the
state $\sigma$. In particular, ${\cal N}_{\Sigma}$ corresponds to the 
majority orientation in the given configuration, and $N =
\sum_\sigma {\cal N}_\sigma$ is the total number of the dimers in
the system. The macroscopic magnetization $m = \langle {\cal M}
\rangle$ is evaluated by averaging over the statistically
independent configurations  obtained as discussed above. We have
also measured the variance, $\langle (\delta {\cal M})^2\rangle$,
$\delta {\cal M} = {\cal M} - m$, which is related to a
generalized susceptibility 
\begin{eqnarray}
\chi = \frac{1}{k_B T} \langle ({\cal M}-m)^2
\rangle \, ,
\end{eqnarray} 
by a fluctuation-dissipation theorem. Furthermore,
we have computed the average energy $U = \langle H \rangle$ (with
$u=U/N$ we will denote the average energy per dimer) as well as
the corresponding variance related to the specific heat per dimer
\begin{eqnarray}
c_N= \frac{1}{N k_B T^2}\langle(H-U)^2\rangle \, .
\end{eqnarray}
 For the other broken
symmetries the fluctuating sublattice magnetizations are defined
to reproduce the definitions given in the previous Section.

In the following, the phase behavior as well as the phase
transitions for the different cases will be illustrated. Since a
mean-field result does not always predict the correct qualitative
behavior, in particular the order of the phase transition, see
Section~\ref{mean-field}, we have put considerable effort to
investigate the nature of the phase boundaries.

\subsubsection{Ferromagnetic system}
By comparison of the free energies of different ordered
configurations we have found that the ground state with the
ferromagnetic broken symmetry is realized for $K>0$ with $M/K>-2$
and for $K<0$ with $M/|K|>5$. Due to the different mechanism of the
ferromagnetic ordering in the two cases, the corresponding phase
transitions to the paramagnetic phase differ in their nature and
characteristics. In the following we will discuss the two regions separately.

\begin{figure}[htbp]
\includegraphics[width=1.0\linewidth]{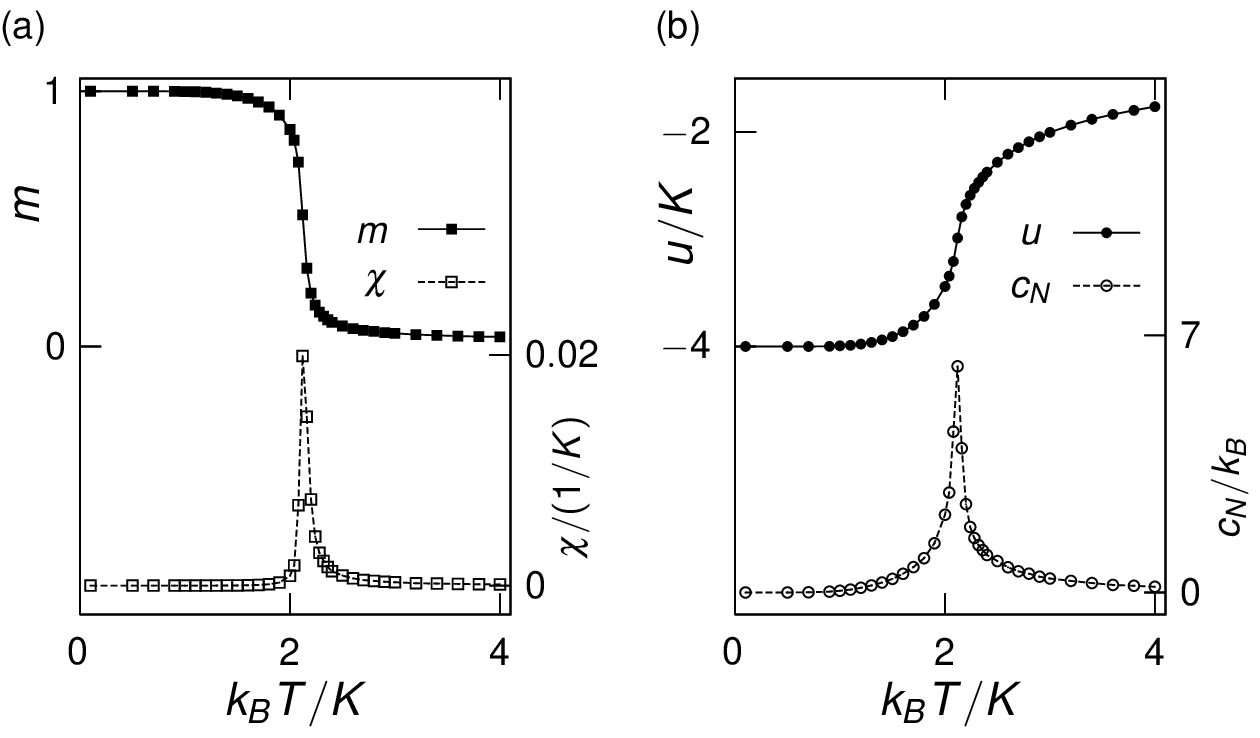}
\vspace{-0.5cm}
\caption{The ferromagnetic system ($K > 0, M/K=1$). Temperature 
dependence (a) of the order parameter and its variance and (b) 
of the energy per dimer and its variance. Both indicate a 
continuous nature of the phase transition.}
\label{fig-5-1}
\end{figure}

(i) For $K>0$ the parallel ordering of dimers is favored by the
$K$-term of the colloidal dimer hamiltonian, Eq.~(\ref{eqn-3-8}),
i.e., by its Potts part. In addition, the remaining $M$-term also
favors this preference and effectively rescales the energy scale
$K$ in the same way as it has resulted from the mean-field
description. In essence, in this regime the colloidal dimer system
can be described by an effective hamiltonian, ${\cal H}_{\rm
eff}=-K_{\rm eff} \sum_{\langle ij \rangle} \delta_{\sigma_i,
\sigma_j}$, where $K_{\rm eff} = K+M/3$. Our Monte Carlo
simulations are in agreement with the known results for such a
hamiltonian. The FM--P phase transition is clearly continuous as
can be inferred from the  pronounced fluctuations of the
magnetization and energy in the vicinity of the transition, see
Fig.~\ref{fig-5-1}. Furthermore, the distribution functions of the
magnetization and energy per dimer exhibit only a single peak in
the whole temperature range under consideration (not shown).

The critical temperature of the FM--P transition closely follows
the ``exact'' value of the Potts hamiltonian,
Eq.~(\ref{eqn-3-10}), provided the exchange coupling $J$ is
replaced by the above mentioned effective energy scale, $K_{\rm
eff}$. Only close to the coexistence region with the HB phase, for
$-1>M/|K|>-2$, the critical temperature deviates from the scaling
law in $K_{\rm eff}$ and the ferromagnetic order melts before the
Potts critical temperature is reached, see Fig.~\ref{fig-4-5}.
Nevertheless, within the numerical precision the nature of the
transition remains unaffected.

(ii) For ${K<0}$ the Potts term of the colloidal dimer
hamiltonian, Eq.~(\ref{eqn-3-8}), disfavors  ferromagnetic order
and anti-ferromagnetic order would be expected. However, for high
values of the energy scale $M$, explicitly for $M/|K|>5$, the term
favoring the alignment of neighboring dimers along the line
connecting them dominates, and the system exhibits a FM symmetry
at low temperatures.

\begin{figure}[htbp]
\includegraphics[width=1.0\linewidth]{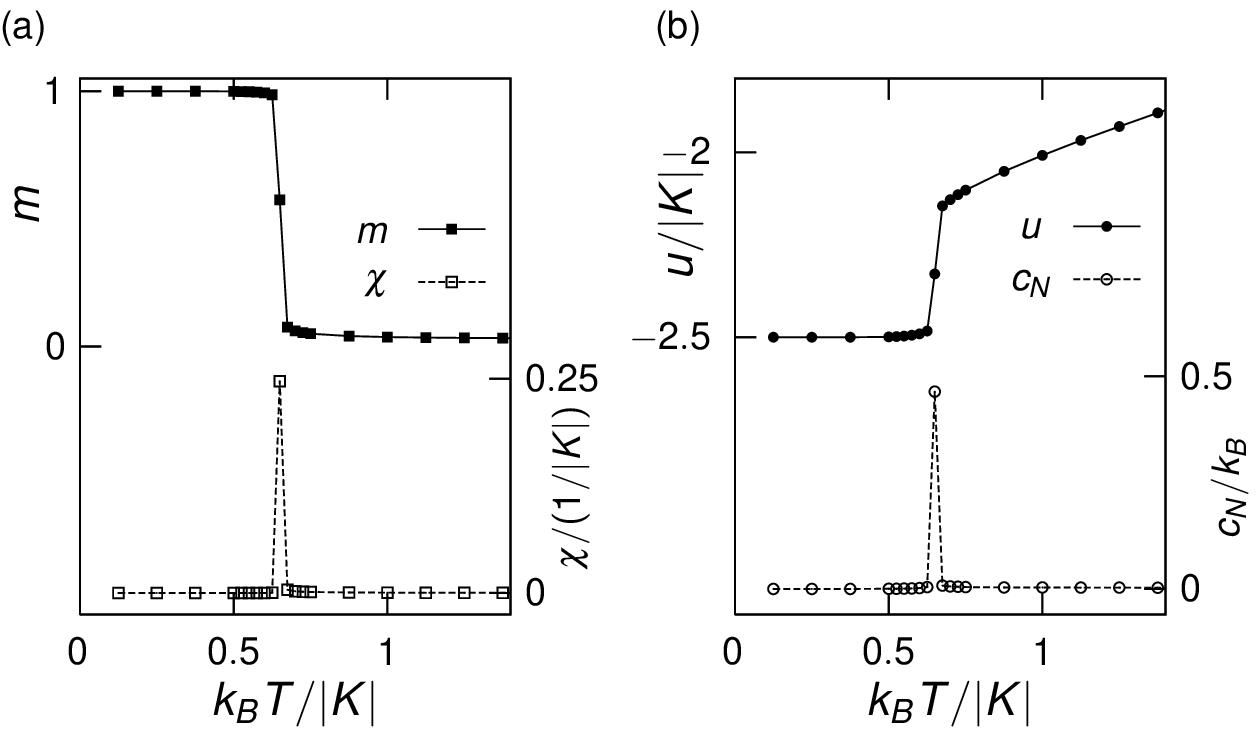}
\vspace{-0.5cm}
\caption{The ferromagnetic system ($K < 0, M/|K|=5.5$). 
Temperature dependence (a) of the order parameter and its variance 
and (b) of the energy per dimer and its variance (right). Both 
indicate a discontinuous nature of the phase transition.}
\label{fig-5-2}
\end{figure}

The critical properties as well as the mechanism of the
transition to the P phase are expected to differ from those of the
Potts hamiltonian. Indeed, the MC simulations of the full
hamiltonian reveal that the phase change is discontinuous.
At the critical point the systems ``jumps'' between the two
degenerate, FM and P, phases, and typical snapshots reveal phase
coexistence regions at the critical point (not shown).
The discontinuous nature of the transition is also evidenced by
the fluctuations of the order parameter and energy, see
Fig.~\ref{fig-5-2}.

The occurrence of metastable phases can be quantified by
determining the distribution functions for the fluctuating values
of the order parameter and energy per dimer~\cite{binder2002}. We
have carefully collected their distributions in the vicinity of
the transition and confirm a discontinuous transition.

\subsubsection{Herring bone phase}
The broken symmetry state with a herring bone pattern
 is realized for $K>0$ and
$M/K<-2$. Within  mean-field theory we have shown that its
ordering is described by two order parameters, i.e., the
magnetization $m$ and the biaxiality parameter $n$. In MC
simulations we have monitored them both.

\begin{figure}[htbp]
\includegraphics[width=1.0\linewidth]{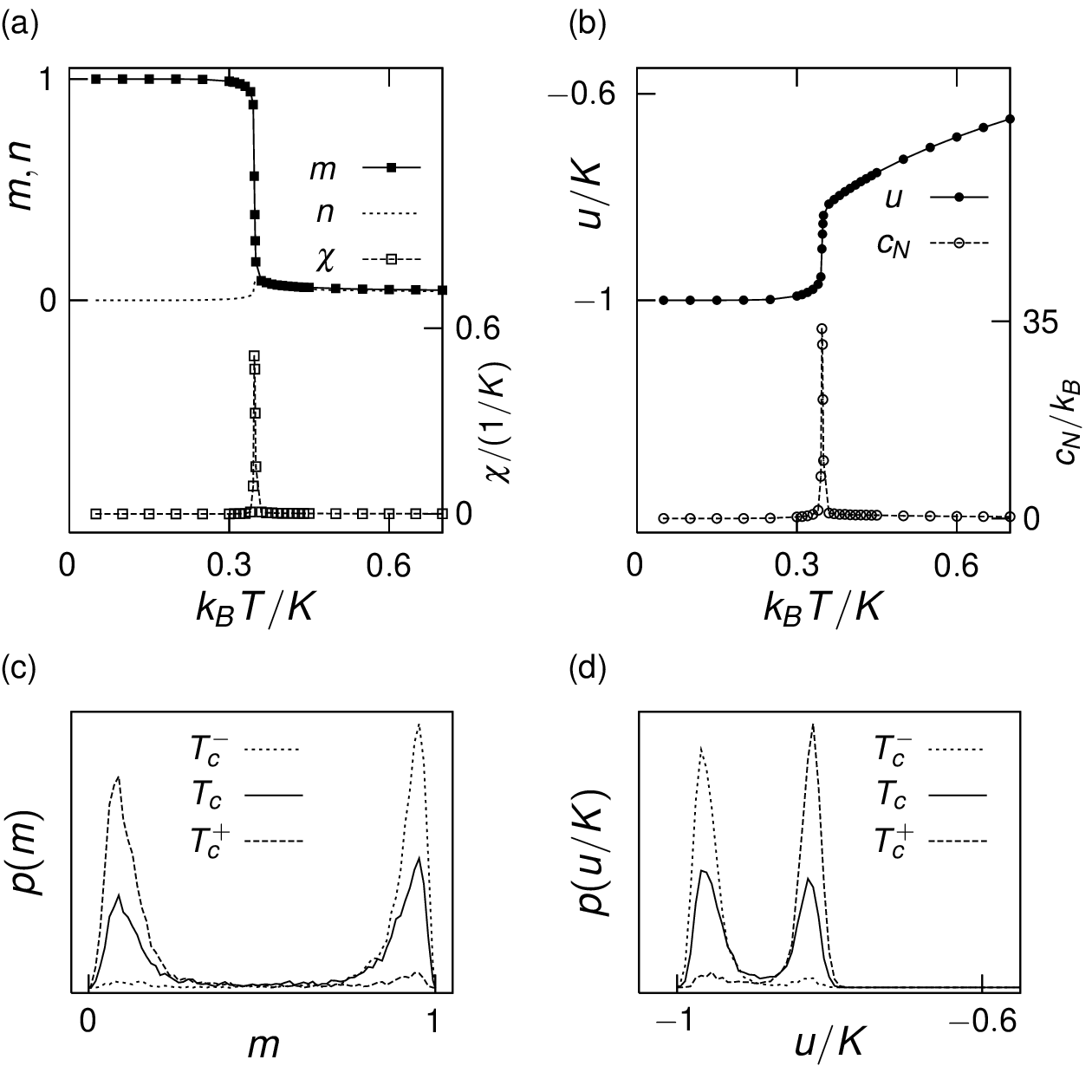}
\vspace{-0.5cm}
\caption{The herring bone system ($K>0, M/K=-2.1$).
Temperature dependence (a) of the order parameter and its variance 
and (b) of the energy per dimer and its variance. The distribution 
functions (c) of the order parameter and (d) energy per dimer. At 
the phase transition, note the bimodal structure characteristic 
for a discontinuous phase transition.}
\label{fig-5-4}
\end{figure}

\begin{figure}[htbp]
\includegraphics[width=1.0\linewidth]{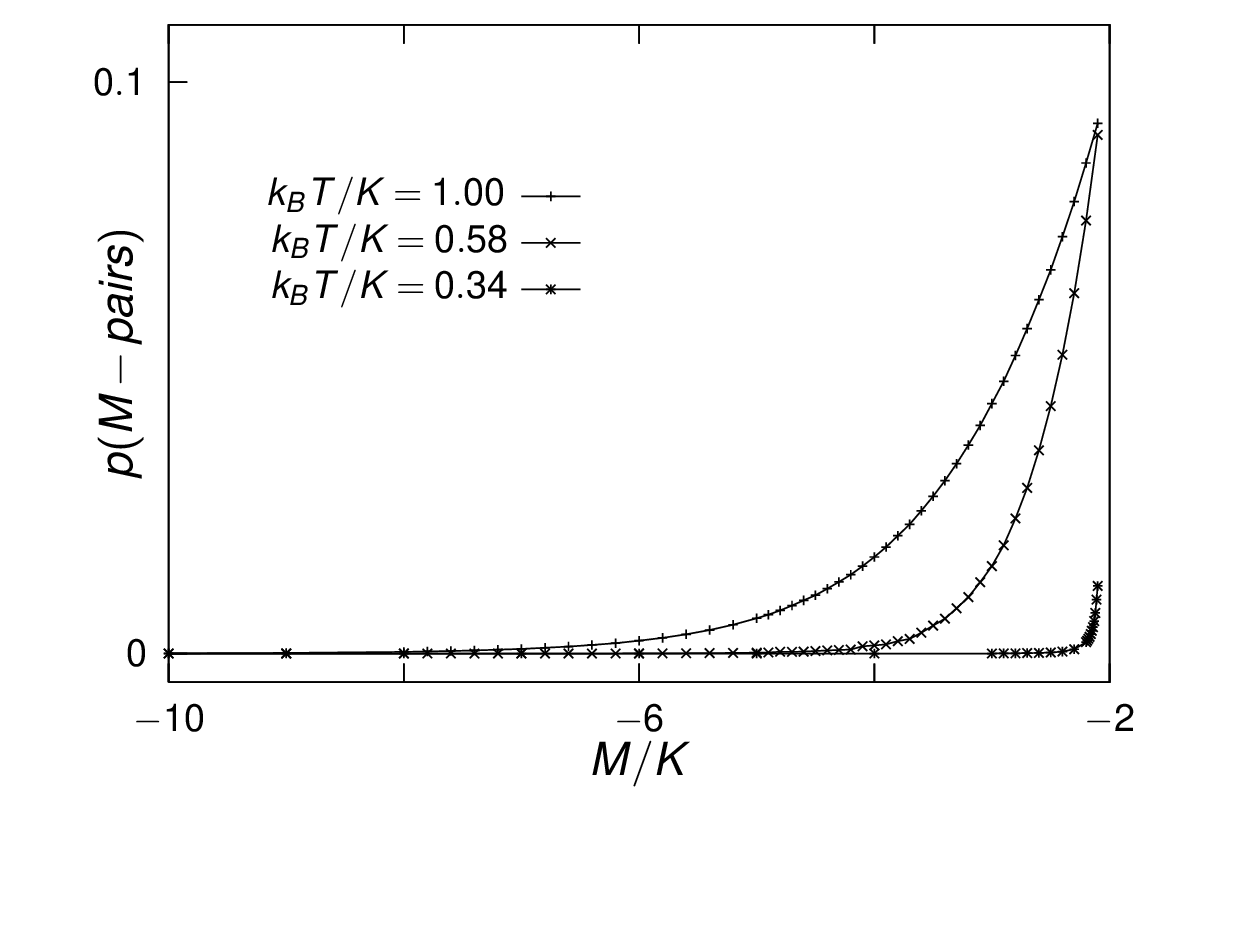}
\vspace{-1.25cm}
\caption{The probability of finding an $M$-pair 
[$p(M$-pairs$)=N_{M{\rm -pairs}}/3N$] in a herring bone system in the 
HB phase ($k_B T/K=0.34$), close to the saturated value of the 
critical temperature ($k_B T/K=0.58$ and $k_B T_c/K \approx 0.6$), 
and well in the P phase ($k_B T/K=1$) as a function of $M/K$. 
Note that the equilibrium probability for $M$-pairs in the normal P 
phase (where the probabilities of all dimer orientations are equal, 
as well as the probabilities of all configurations of dimer pairs) 
is 1/9.}
\label{fig-5-6}
\end{figure}

Qualitatively, the MC temperature dependences of $m$ and $n$ are
the same as the ones determined within the mean-field description.
We find a discontinuous phase change  for $M/K \lesssim -2$,
identified by jumps of the magnetization and energy per dimer at
the critical temperature. The jump of the biaxiality parameter is
masked by rounding due to the finite size of the system. However,
we find a sharp increase upon approaching the critical point from
below, similar to the one found within the mean-field description,
see Fig.~\ref{fig-5-4}. For increasing negative values of $M/K$
the latent heat decreases, so that below $M/K=-4$ we cannot
distinguish between a weakly discontinuous and a continuous
transitions. Correspondingly, it is difficult to clearly
corroborate the existence of the tricritical point predicted by
mean-field at $M/K =-7/3$, let alone to determine its position.
Monitoring the probability distribution of the order parameters
suggests that indeed there is a change of the nature of the phase
transition in the vicinity of the mean-field prediction.

It is known that quantitatively, the effect of fluctuations
decreases the critical temperature  obtained within mean-field. As
it can be clearly seen in Fig.~\ref{fig-4-5}, for the HB--P
transition this appears  to be not the only effect. If in the
interval $-2>M/K\gtrsim -3$ there is still some evidence of the
scaling law similar to the one found within the mean-field
description, this practically vanishes for $M/K<-3$. Finally,
for $M/K<-4$ the critical temperature ``saturates'' at a value
$k_B T_c/K \approx 0.6$, i.e., it does not depend any more  on
the energy ratio $M/K$.

To rationalize this finding, let us recall the expression for the
colloidal dimer hamiltonian~(\ref{eqn-3-8}). The first, $K$-term,
favors for $K>0$  parallel alignment of dimers. On the other hand,
the $M$-term with a large negative value of $M$ strongly
suppresses  parallel alignment in which the two dimers face each
other -- to be referred to as an  $M$-pair,
Fig.~\ref{fig-3-1}(b). As a consequence, the probability of
finding  $M$-pairs in either of the two phases, HB or P, is
rapidly decreasing with  increasing negative values of $M/K$.
The number of $M$-pairs as a function of $M/K$ is shown in
Fig.~\ref{fig-5-6} for three paths of a constant temperature,
viz. in the HB phase, in the P phase, and at the intermediate
temperature $k_B T/K=0.58$ for which the system traverses the phase 
boundary from HB to P at $M/K \approx -3.7$.
For large
enough negative $M$ the role of the $M$-term is only to reduce the
number of degrees of freedom in the system. Then the corresponding
critical behavior depends solely on the relative magnitude of the
exchange energy $K$ with respect to the thermal energy $k_B T$. In
the ($k_B T/|K|$,$M/|K|$) phase diagram such a phase boundary is
represented by a straight vertical line. Within  mean-field
the reduction of the number of degrees of freedom for
the configurations of dimer pairs is not properly captured. Thus the role
of the $M$-term is as in all the other cases reduced to
effectively rescale the exchange energy $K$.

\subsubsection{Anti-ferromagnetic phase}
The ground state of the anti-ferromagnetically ordered phase occurs in
colloidal dimer systems with $K<0$, $M/|K|<1$. As discussed
in Sec.~\ref{mean-field}, its order can be
described by a single order parameter -- a sublattice
magnetization $m$, where the system is divided into three
triangular sublattices. 

\begin{figure}[htbp]
\includegraphics[width=1.0\linewidth]{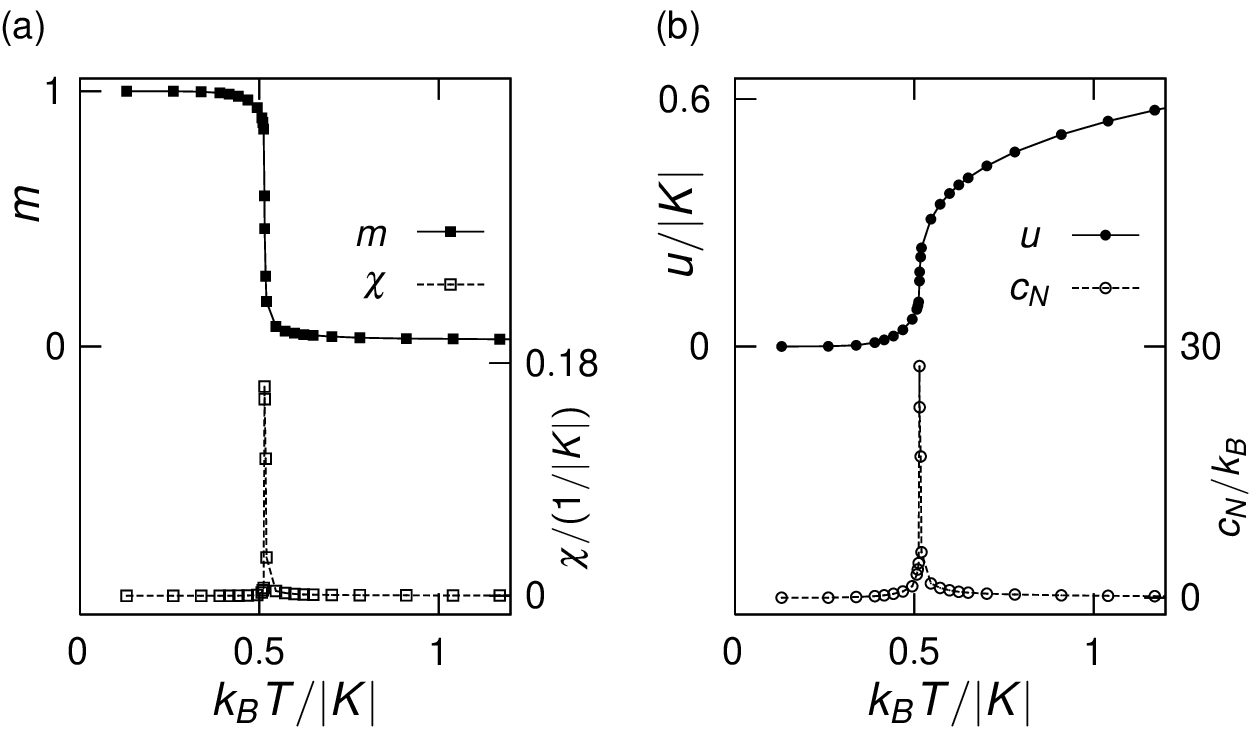}
\vspace{-0.5cm}
\caption{The anti-ferromagnetic system ($K < 0, M/|K|=0.4$). 
Temperature dependence (a) of the order parameter and its variance 
and (b) of the energy per dimer and its variance.}
\label{fig-5-7}
\end{figure}

The AM--P phase transition is discontinuous in the whole range of
the stable AM phase. In contradiction to the {\em universal} 
jump of the latent
heat and heat capacity determined by the mean-field approach, the
MC results show slightly decreasing strength of the discontinuity
of the transition when $M/|K|$ is increased towards $1$. At
the special point, $M=0$, where the colloidal dimer
hamiltonian reduces to the Potts model, our results match the ones
known for the Potts anti-ferromagnet~\cite{saito1982,wu1982}. Applying 
again the empirical shift of the mean-field analysis to match the 
anti-ferromagnetic Potts point, gives a satisfactory description of
the phase boundary for $M/|K| \gtrsim 1$, see Fig.~\ref{fig-4-5}. 
For highly negative
$M/|K|$ the critical temperature saturates as is the case of
the HB--P transition. The saturation again originates in the
suppression of  $M$-pairs  as  already discussed above. 

\subsubsection{Japanese 6 in 1 system}
\begin{figure}[htbp]
\includegraphics[width=1.0\linewidth]{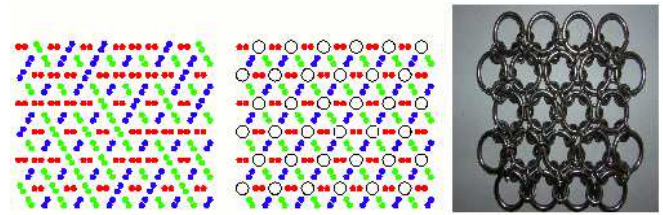}
\caption{The Japanese 6 in 1 system. Left: A snapshot of one of its 
ground states. Middle: Schematic representation of the J6/1 
configuration. Here the circles indicate the degeneracy of the 
dimer orientation on the disordered sublattice. Right: A picture of 
chain maille using the Japanese 6 in 1 weaving 
pattern~\cite{asa-no-ha-gusari}.}
\label{fig-5-8}
\end{figure}

For $K<0$, squeezed between the FM and AM phases one observes the
new J6/1 phase. Its highly degenerate ground state is stable for
$1<M / |K|<5$. The degeneracy of the J6/1 phase originates in
the disordered nature of one of its sublattices. A typical
snapshot of a ground state in the J6/1 phase is presented in
Fig.~\ref{fig-5-8}. In addition, a schematic representation and
a picture of a chain maille using the Japanese 6 in 1 weaving
pattern~\cite{asa-no-ha-gusari} from which we have borrowed the
name are plotted.

\begin{figure}[htbp]
\includegraphics[width=1.0\linewidth]{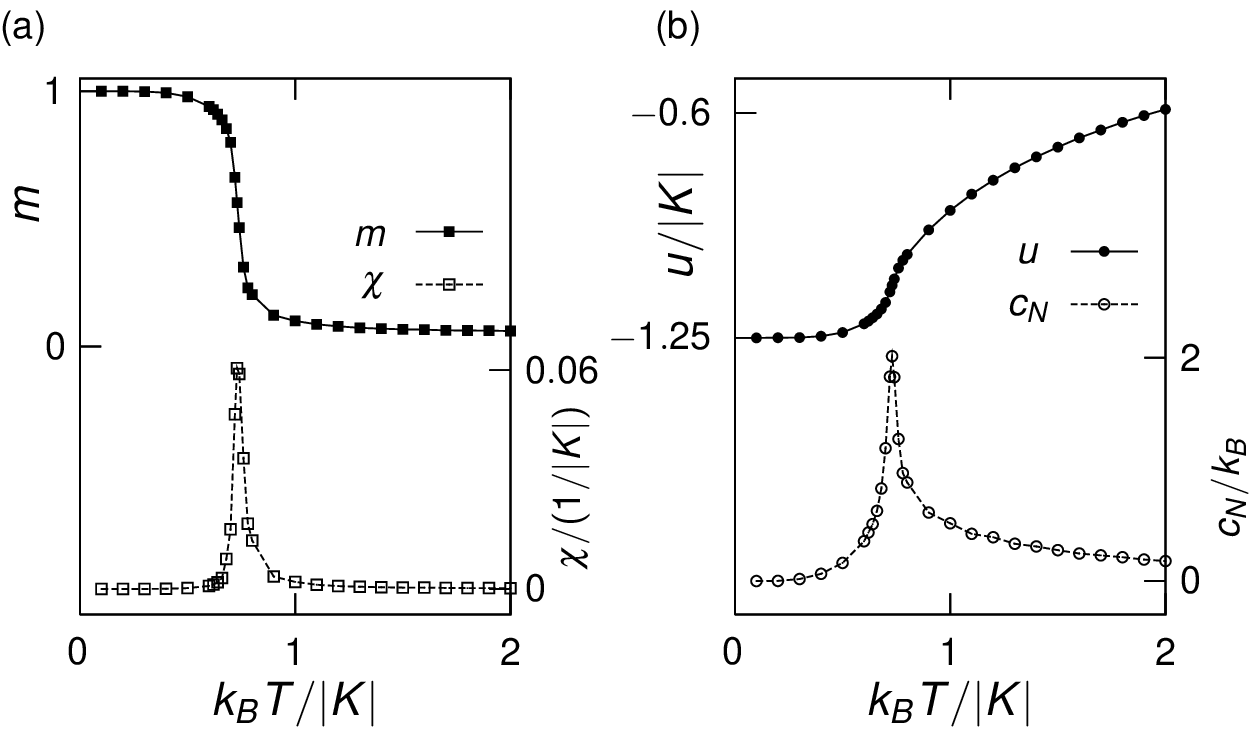}
\vspace{-0.5cm}
\caption{The Japanese 6 in 1 system ($K<0, M/|K|=3.5$). 
Temperature dependence (a) of the (sublattice) order parameter and 
its variance and (b) of the energy per dimer and its variance (right). 
Both indicate a continuous nature of the phase transition.}
\label{fig-5-10}
\end{figure}

\begin{figure}[htbp]
\includegraphics[width=1.0\linewidth]{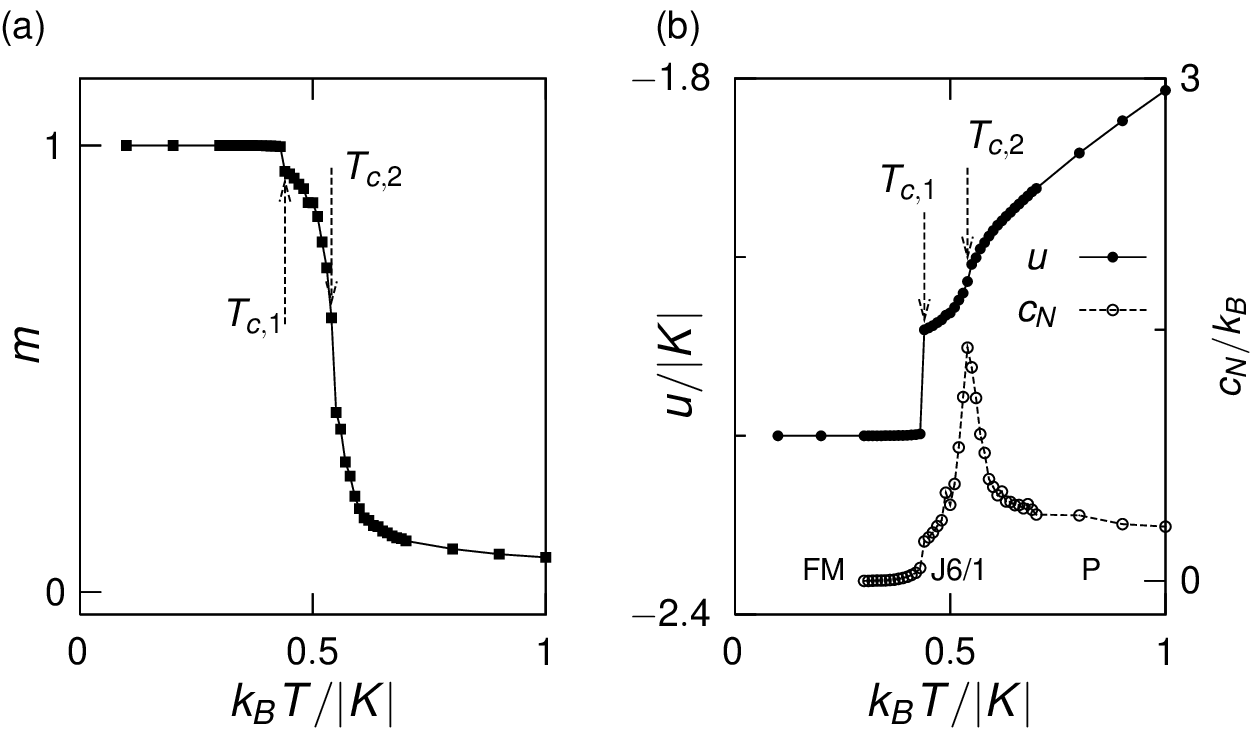}
\vspace{-0.5cm}
\caption{ The system ($K<0, M/|K|=5.2$)
exhibits  a sequence of phase transitions FM--J6/1--P upon 
increasing the temperature. The FM--J6/1 
structural transition takes place at $T_{c,1}$ and is 
discontinuous whereas there is no jump in the order parameter and 
energy per dimer at the J6/1--P transition, which takes place at 
$T_{c,2}$.}
\label{fig-5-11}
\end{figure}

The MC simulations corroborate the division of the J6/1 system
into four triangular sublattices. Three of them are equivalent and
ordered with mutually different preferred dimer orientations for
temperatures below the critical one. The remaining sublattice exhibits no
preferred orientation for all temperatures. The magnetization parameter 
is defined as the average of the three non-zero sublattice magnetizations,
${\cal M}=\sum_\Gamma {\cal M}_\Gamma/3$. In the simulation we
have monitored both, the magnetization of the ordered 
as well disordered sublattice. Upon
increasing the temperature from the ground state, the order in the
ordered sublattices of the J6/1 phase melts and the system
exhibits a continuous phase transition to the P phase. The
corresponding phase boundary is represented by a curved,
non-monotonic line, see Fig.~\ref{fig-4-5}. Close to the critical
point the fluctuations of the order parameter are strongly
enhanced and the system can ``switch'' between the four classes of
configurations, i.e., characterized by a translation of the
disordered sublattice. Graphs of the typical temperature
dependences of the sublattice magnetization and energy per dimer
are depicted in Fig.~\ref{fig-5-10}.

The high residual entropy stabilizes the J6/1 with respect to the neighboring
FM or AM phase upon increasing temperature. Correspondingly, the 
parameter region where the J6/1 phase is stable 
expands upon heating. For parameters 
$5 < M/|K| \lesssim 5.3$ or $0.8 \lesssim M/|K| < 1$ the melting 
of the FM or AM proceeds in  two steps via 
a discontinuous structural
transition to the J6/1 phase followed by another transition to the
paramagnetic state. A typical temperature variation of the
energy per dimer and the corresponding heat capacity for the case
of a FM--J6/1--P sequence is presented in Fig.~\ref{fig-5-11}. The
AM--J6/1--P phase sequence is characterized by similar graphs.

\section{Conclusions}
We have shown that soft matter colloidal systems exhibit peculiar
phase behavior once exposed to external laser fields. The
laser-colloid interaction favors configurations where the
particles are attracted to the spots of high light intensities. By
a subtle balance of the compression of groups of two or three
particles at such spots and their mutual repulsion composite
objects are formed, such as dimers and trimers. The relevant
low-energy degrees of freedom are then the discrete orientational
configurations with an excitation spectrum at the thermal scale.

The introduction of an effective hamiltonian allows to map the
phase behavior to spin-like systems. In particular, we have
rationalized the case of trimers on a triangular
lattice~\cite{brunner2002}, which has motivated this work, in
terms of a two-dimensional Ising model. A comparison with
experimental data corroborates the approach of using composite
objects and validates our theory.

We have suggested a new experimental setup  where dimers play the
role of the composite objects. It is shown that a rich phase
behavior emerges; in terms of a ``spin language'' a ferromagnetic,
an  anti-ferromagnetic, a herring-bone, and an exotic Japanese 6 in
1 phase have been determined. We have mapped the phase diagram
using a variational mean-field theory and substantiated our
findings by Monte Carlo simulations.

It is interesting to note that these colloidal molecular crystals
have a counterpart on the atomic scale, viz. adsorption of
molecules on atomic surfaces. For instance, nitrogen on
graphite~\cite{chung1977,diehl1982} at suitable densities orders
in a herring-bone structure similar to the one of colloidal
dimers, which lead Sluckin~\cite{sluckin1988}
to construct a generalized Potts model closely along the line of
reasoning presented here. Allen and Armitstead~\cite{allen1989}
presented Monte Carlo simulations close to the transition from
ferromagnetically ordered nitrogen adsorbats to the experimentally
observed herring-bone structure. Using solely symmetry arguments,
Vollmayr~\cite{vollmayr1992} discussed the possible ground states
if the interaction between two molecules depends on the  bond
orientation and explored regions of suppressed relaxations in an
{\em a priori}  unfrustrated system.

It appears promising to extend the concept of composite objects to
different symmetries of the underlying lattice, e.g. square or
rectangular unit cells, and to non-integer filling factors. The
corresponding phase diagrams are expected to include a variety of
phases and phase transition.

\section*{Acknowledgments}
We thank C. Bechinger for stimulating discussions and acknowledge 
helpful comments by C. Reichhardt. A.\v{S}.
acknowledges the support by the Alexander von Humboldt Foundation.
T.F. gratefully acknowledges the kind hospitality of the
University of Ljubljana where part of this work has been finished.

\appendix
\section{Relation between energy scales and experimental control
parameters}\label{appendix-1}
In this Appendix we show how the parameters of the trimer
hamiltonian can be expressed in terms of the laser intensity, the
screening length of the colloid-colloid interaction, and the
effective charge of the colloids. The case of dimers is then a
straightforward modification and we merely state the result at the
end of this Appendix.

The formation of composite objects, i.e., trimers, as well as
their mutual interaction is due to a subtle interplay of the
external laser potential and the screened Coulomb colloid-colloid
repulsion. The symmetry of the external potential favors
equilateral triangles aligned along one of the two distinguished
lattice directions. The potential squeezes the constituents of the
trimer closer to each other. The forces of the external potential
on a trimer are balanced by the internal constituent Coulomb
repulsion. To lowest order the effect of the neighboring trimers
can be ignored since in the regime of interest the trimers are
small compared to the lattice constant. Furthermore the Coulomb
interaction is exponentially suppressed for the case of screening
lengths smaller than the lattice constants. The thermal pressure
is negligible compared to the {\em binding energy}.

To this order of approximation the ground state of the system has
a two-fold degeneracy per lattice site. Introducing the residual
interaction of the trimers, the energy levels are split on a scale
comparable with the thermal one.

Let us discuss the interactions of the trimer constituents in
detail. First the dielectric contrast at optical frequencies of
the colloidal particles with respect to the solvent gives rise to
a coupling to the laser intensity. This interaction can be
expressed in terms of an effective potential. For the experimental
setup under consideration the total electric field consists of a
coherent superposition of three laser beams
\begin{eqnarray}
    {\bf E}({\bf r},t) = {\bf E}_0 e^{-i\omega t}
        \left[ \, {\rm e}^{i {\bf k}_1 \cdot {\bf r}}
        + \, {\rm e}^{i {\bf k}_2 \cdot {\bf r}}
        + \, {\rm e}^{i {\bf k}_3 \cdot {\bf r}}\right]
        + \mbox{c.c.} ,
\label{eqn-1-1}
\end{eqnarray}
where the respective in-plane projections of wave vectors
${\bf k}$ constitute an equilateral triangle. The total intensity
consists of a rapidly varying component in time, which averages to
zero. The remaining time-independent component gives rise to a
constant, which we also discard, and the interesting part,
characterized by a spatial modulation,
\begin{eqnarray}
    I({\bf r}) = 4 |E_0|^2 [\cos({\bf G}_1 \cdot {\bf r})
        + \cos({\bf G}_2 \cdot {\bf r})
        + \cos({\bf G}_3 \cdot {\bf r})] ,
\label{eqn-1-2}
\end{eqnarray}
where ${\bf G}_1={\bf k}_2-{\bf k}_3$,
${\bf G}_2={\bf k}_3-{\bf k}_1$, and
${\bf G}_3={\bf k}_1-{\bf k}_2$ again form an equilateral
triangle.

The periodic potential exhibits the symmetry of a triangular
lattice with ${\bf G}_i$'s being the wave vectors of the
reciprocal lattice. In the following we choose the coordinate
system in which
\begin{equation}
    {\bf G}_1 =  G (0,-1)  \, , \quad
    {\bf G}_2 =
        G (\textstyle{\sqrt{3} \over 2},\textstyle{1 \over 2})  \, , \quad
    {\bf G}_3 =
        G (-\textstyle{\sqrt{3} \over 2},\textstyle{1 \over 2})\, .
\label{eqn-1-3}
\end{equation}
It is also convenient to introduce the spacial lattice vectors
\begin{eqnarray}
    {\bf a}_1 = a (1,0)  \, , \quad
    {\bf a}_2 =
        a (\textstyle{1 \over 2},\textstyle{\sqrt{3} \over 2})  \, , \quad
    {\bf a}_3 =
        a (-\textstyle{1 \over 2},\textstyle{\sqrt{3} \over 2}) \, .
\label{eqn-1-4}
\end{eqnarray}
where $a=4\pi/(G\sqrt{3})$ is the lattice constant of the external
potential. In this parameterization and with respect to the
lattice sites the relative positions of the colloidal particles
within a trimer with orientation $S\in\{+1,-1\}$ are parameterized
by the single dimensionless distance parameter $\rho$:
\begin{eqnarray}
    {\bf R}_1 = S \, \rho {\bf a}_1\, , \quad
    {\bf R}_2 = S \, \rho {\bf a}_2 \, , \quad
    {\bf R}_3 = S \, \rho {\bf a}_3 .
\label{eqn-1-5}
\end{eqnarray}

The effective interaction of the colloidal particles with the
external electric potential is proportional to the laser intensity
and the dielectric contrast at optical frequency, i.e.,
\begin{equation}
    \varphi_D ({\bf r}) = - V_0 \, [I({\bf r}) / 4 |E_0|^2] ,
\label{eqn-1-2a}
\end{equation}
where $V_0 > 0$ equals $V_0$ used for the parametrization of the 
external potential in experiment of Ref.~\cite{brunner2002}. 

The strong mutual repulsion of the three constituents is mediated
via a screened Coulomb interaction
\begin{equation}
    \varphi_C (r) = {q^2 \over r} \, {\rm e}^{- \kappa r} ,
\label{eqn-1-6}
\end{equation}
where $\kappa$ is the inverse Debye length, $r$ is the center to
center distance, and
$q^2=[(Z^\ast e_0)^2/(4\pi\epsilon_0\epsilon_r)]\
\, {\rm e}^{\kappa R_s}/(1+\kappa R_s)$ with $Z^\ast e_0$ being the charge
on the surface of a single particle, $\epsilon_r$ the dielectric
contrast of water, and $R_s$ the radius of a colloidal
particle~\cite{derjaguin1941,vervey1948}. Introducing dimensionless
parameters $\rho=r/a$ and $k=\kappa a$ the screened Coulomb repulsion
scales as $\varphi_C (\rho) = {\cal E}_0 \exp{(-k\rho)}/\rho$, where
${\cal E}_0 = q^2/a$.

The internal potential energy  of a single trimer,
$E_C = \sum_{i<j} \varphi_C(\rho_{ij})$ and $\rho_{ij}$ the distance of
two colloids within the trimer,
\begin{equation}
    E_C/{\cal E}_0 = \, {\sqrt{3} \over \rho} \, {\rm e}^{- k \rho \sqrt{3}} ,
\label{eqn-1-7}
\end{equation}
is competing with the compression of the external dielectric
potential, $E_D = \sum_i \varphi_D(\rho_i)$ and $\rho_i$ the distance 
of a colloid from the center of mass of the trimer, 
\begin{equation}
    E_D/{\cal E}_0 = -3 v_0 (1 + 2 \cos{2 \pi \rho}) ,
\label{eqn-1-8}
\end{equation}
where $v_0 = V_0/{\cal E}_0$. 

\begin{figure}[htbp]
\includegraphics[width=1.0\linewidth]{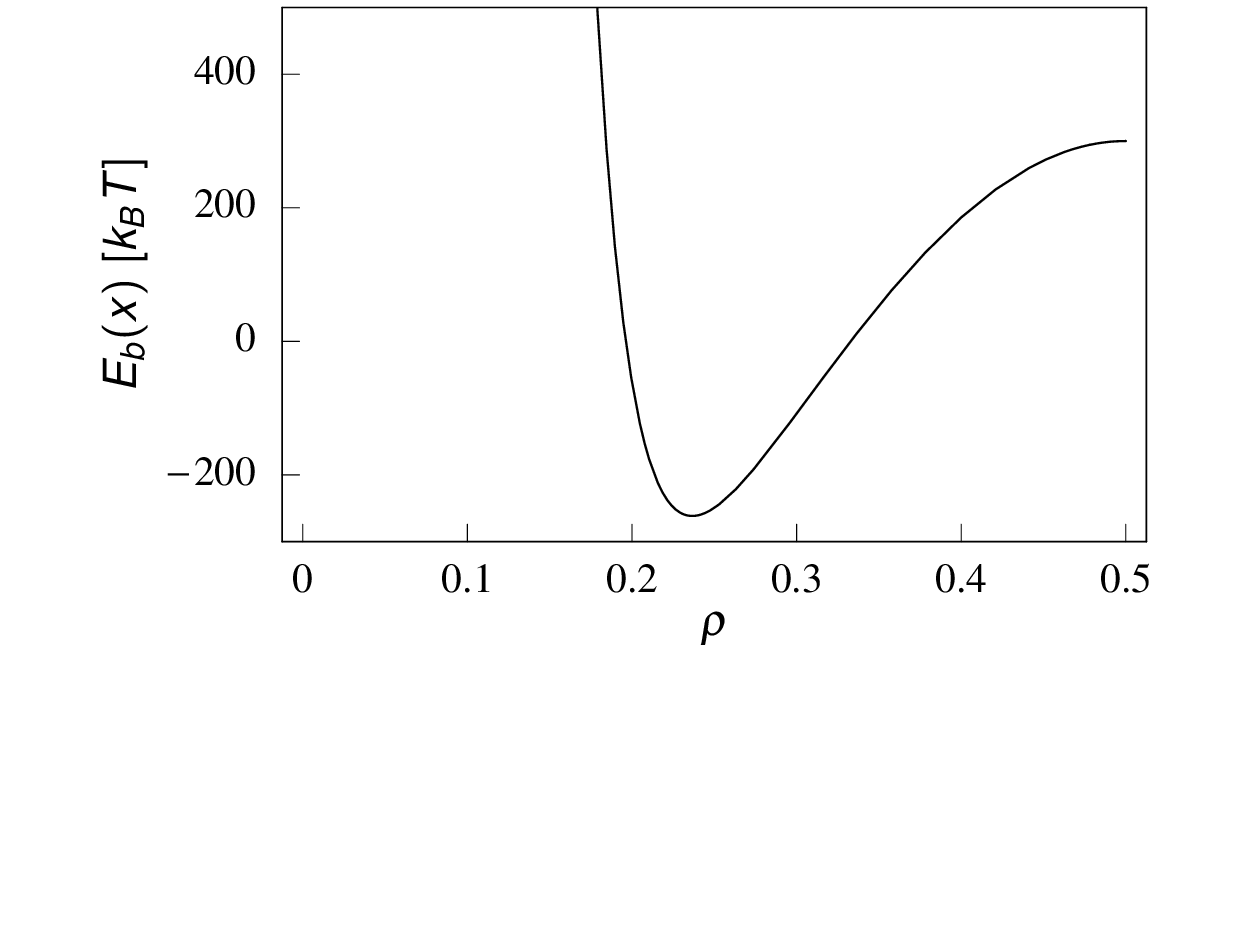}
\vspace{-2.25cm}
\caption{Dependence of the binding energy of a trimer on its size. 
The equilibrium trimer extension is determined by the minimum of 
the binding energy, $\rho\leq 1/3$.}
\label{fig-1-1}
\end{figure}

The size of a trimer is determined by minimizing the total
potential energy, $E_C + E_H$, see Fig.~\ref{fig-1-1}, with
respect to $\rho$. The minimum defines the {\em binding energy}
$E_b$, which for typical experimental parameters exceeds the
thermal scale by approximately two orders of magnitude --
$E_b\sim 10^2~k_B T$, for $\kappa=570$~nm, $a=11.5~\mu$m,
${\cal E}_0=10^5~k_B T$, and $V_0=60-110~k_B T$.
The separation of scales justifies the idea
to treat the assembly of three colloidal particles corresponding
to a lattice site as a rigid composite object.

In the picture of composite objects the trimer-trimer interaction
enters as residual interaction expected to be much smaller than
the binding energy. The trimer-trimer interaction lifts the
orientational degeneracy and is the origin of the ordering
phenomena under consideration. Since the Coulomb interaction is
strongly screened it is sufficient to consider only neighboring
lattice sites. For the three different geometrical arrangements of
the trimers, see Fig.~\ref{fig-2-1}, the interaction energy is
determined by the sum of the nine direct colloid-colloid
contributions of the rigid triangles,
\begin{eqnarray}
 E_\mu = {\cal E}_0 \sum_{s,t=1}^3 {1 \over \rho_\mu[s,t]} \exp(-k \rho_\mu[s,t]) .
\label{eqn-1-9}
\end{eqnarray}
Explicitly, elementary geometrical considerations yield for the
tables of the distances in dimensionless units
{\small
\begin{eqnarray}
\rho_1 &= & \left[ \begin{array}{ccc}
         1+2 \rho_0 & \sqrt{1+\rho_0+\rho_0^2} & \sqrt{1+\rho_0+\rho_0^2}   \\
          \sqrt{1+\rho_0+\rho_0^2} & \sqrt{1-2\rho_0+4\rho_0^2} & 1-\rho_0 \\
          \sqrt{1+\rho_0+\rho_0^2} &   1-\rho_0 & \sqrt{1-2\rho_0+4\rho_0^2}
    \end{array} \right] , \nonumber \\
\rho_2 &= & \left[ \begin{array}{ccc}
  1 & \sqrt{1-3\rho_0+3\rho_0^2} & \sqrt{1-3\rho_0+3\rho_0^2} \\
 \sqrt{1+3\rho_0+3\rho_0^2} & 1 & \sqrt{1+3\rho_0^2} \\
 \sqrt{1+3\rho_0+3\rho_0^2} &  \sqrt{1+3\rho_0^2} & 1
\end{array}\right] ,
 \nonumber \\
\rho_3 &= & \left[ \begin{array}{ccc}
  1-2\rho_0 &  \sqrt{1-\rho_0+\rho_0^2} & \sqrt{1-\rho_0+\rho_0^2}  \\
\sqrt{1-\rho_0+\rho_0^2} &  \sqrt{1+2\rho_0+4\rho_0^2} &  1+\rho_0 \\ \sqrt{1-\rho_0+\rho_0^2}
& 1+\rho_0 & \sqrt{1+2\rho_0+4\rho_0^2}
\end{array}\right] , 
\label{eqn-1-10}
\end{eqnarray}
}where $\rho_0$ is the equilibrium size of the trimer. For the
experimental parameters used above, one obtains $E_1\sim 0.01~k_B
T, E_2\sim 0.1~k_B T, E_3\sim 1~k_B T$ and, thus, $J>0$ [see
Eq.~(\ref{eqn-2-4})], which shows that phase transitions are to be
expected in the realistic experimental setups.

For the case of dimers the chain of arguments is along the same
lines as above, and one finds for the  dimer-dimer interactions in
the four different geometrical arrangements exhibited in
Fig.~\ref{fig-3-1}~(b),
\begin{eqnarray}
 E_\mu = {\cal E}_0 \sum_{s,t=1}^2 {1 \over \rho_\mu[s,t]} \exp(-k \rho_\mu[s,t]) ,
\label{eqn-1-11}
\end{eqnarray}
where
\begin{eqnarray}
\rho_1 &= & \left[ \begin{array}{cc}
         1 & \sqrt{1+2\rho_0+4\rho_0^2}  \\
          \sqrt{1-2\rho_0+4\rho_0^2} & 1
    \end{array} \right] , \nonumber \\
\rho_2 &= & \left[ \begin{array}{cc}
  1+\rho_0 & \sqrt{1+3\rho_0^2}  \\
\sqrt{1+3\rho_0^2} & 1-\rho_0
\end{array}\right] , 
 \nonumber \\
\rho_3 &= & \left[ \begin{array}{cc}
  \sqrt{1-3\rho_0+3\rho_0^2} & \sqrt{1+3\rho_0+3\rho_0^2}  \\
\sqrt{1-\rho_0+\rho_0^2} & \sqrt{1+\rho_0+\rho_0^2}
\end{array}\right] , 
 \nonumber \\
\rho_4 &= & \left[ \begin{array}{cc}
  1 & 1+ 2\rho_0  \\
1-2\rho_0 & 1
\end{array}\right] .
\label{eqn-1-12}
\end{eqnarray}
For the experimental parameters, as used in the experiment with
trimers, one obtains $E_1 \sim 10^{-1}~k_B T, E_2 \sim 1~k_B T, 
E_3 \sim 10~k_B T, E_4 \sim 10^2~k_B T$, or equivalently, $K > 0$ 
and $M \ll 0$
which sets the colloidal dimer system in the regime of the herring
bone ground state. In order to explore the other regions of the
phase diagram one would have to change the colloid-colloid
interaction, for example, paramagnetic particles could be used.

Small residual thermal fluctuations of the composite objects,
e.g., stretching modes or small oscillations around the minimum of
a gross orientational state, can be absorbed in temperature
dependent nearest-neighbor interaction energies. For example, one
can use a perturbative approach, as has been done for the elastic
constants in the case of colloidal suspensions in one dimensional
troughs~\cite{frey1999}.

\end{document}